\pdfoutput=1
\documentclass[%
preprint,
superscriptaddress,
aip,
]{revtex4-2}

\usepackage{graphicx}
\usepackage{dcolumn}
\usepackage{bm}
\usepackage{amsmath,amssymb} 
\usepackage{subcaption}
\usepackage{booktabs}
\usepackage{tabularx}
\usepackage[dvipsnames]{xcolor}
\usepackage[bottom]{footmisc}
\usepackage[final]{changes}
\usepackage{subcaption}
\setaddedmarkup{{\color{Red}#1}}
\setdeletedmarkup{{\color{Red}\sout{#1}}}
\usepackage{hyperref}
\hypersetup{
    colorlinks=true,
    linkcolor=blue,
    citecolor=blue,
    urlcolor=blue
}
\begin{document}


\title{ Asymptotic state of nonlinear Landau damping in one-dimensional plasma}

\author{Yifei Ouyang}
\affiliation{ School of Physics, Huazhong University of Science and Technology, Wuhan,  430074, China}
\author{ Ping Zhu}
\email{zhup@hust.edu.cn}
\affiliation{ State Key Laboratory of Advanced Electromagnetic  Technology, International Joint
Research Laboratory of Magnetic Confinement Fusion and Plasma Physics, School of Electrical and
Electronic Engineering, Huazhong University of Science and Technology, Wuhan, 430074, China}
\affiliation{Department of Nuclear Engineering and Engineering Physics, University of Wisconsin-Madison, Madison, Wisconsin 53706, USA}
\author{Chung-Sang Ng}
\affiliation{Geophysical Institute, University of Alaska Fairbanks, Fairbanks, Alaska 99775, USA}

\date{\today}
\begin{abstract}
  In this work, the asymptotic state of nonlinear Landau damping in one-dimensional plasma has
been examined using a quasi-linear model and a second-order symplectic integrator. The dispersion
relation of the plateau distribution function \added{for} the steady-state solution of the quasi-linear mode
is extended to the complex plane and compared with the nonlinear simulation.  
 We determine that the asymptotic state of the collisionless plasma is a multi-wave BGK structure. This structure is 
 characterized by multiple vortices in phase space, which correspond to distinct peaks in the frequency-wavenumber (\(\omega,k\))
 spectrum of the electric field.
\end{abstract}

\maketitle
\section{Introduction}\label{section:1}
In his pioneering 1946 paper, \cite{Landau1946}  Landau demonstrated 
that infinitesimal amplitude plasma waves in collisionless plasma can 
be damped, which originates from the resonance 
between the wave and particles with velocity near the phase velocity of the wave. The damping rate is proportional to 
the slope of the initial distribution function at the phase velocity. 
This process has been repeatedly verified by experiments  \cite{malmberg1964collisionless}
and numerical simulations.  \cite{cheng1976integration,maekaku2024time}

Linear Landau damping is widely observed in small-amplitude cases  \cite{zhou2001numerical}
 (except for some specific  initial distribution functions   \cite{belmont2008existence,chust2009landau}).
 However, when a strong perturbation occurs, nonlinear Landau damping 
 arises: particles near the phase velocity are trapped, forming 
 vortex-like structures known as BGK (Bernstein-Greene-Kruskal) modes.  \cite{bernstein1957exact}
 The explanation for nonlinear Landau damping can be traced back to O'Neil's 
calculations on the damping of constant-amplitude perturbations.  \cite{o1965collisionless}
In his results, due to the presence of nonlinear effects (trapped particles), 
the damping rate in the asymptotic state tends to zero over time.

Around 1990, Dorning and Holloway obtained small-amplitude self-consistent 
solutions for single-wave BGK mode.  \cite{holloway1991undamped,holloway1989undamped}
For small wave numbers, they found an additional solution besides the 
Langmuir wave, known as the electron acoustic wave (EAW). The velocity 
distribution function near the phase velocity becomes flat due to trapped 
particles. Based on this, Buchanan and Dorning further constructed 
approximate solutions for multi-wave BGK modes using the first-order local 
invariants obtained from the Hamiltonian Lie perturbation method.  \cite{buchanan1993superposition,buchanan1994near}
Subsequent simulations have confirmed their conclusions to some extent.
  \cite{demeio1991numerical}

Further research has been conducted on both the 
phase-space holes  and 
the  EAWs in collisionless plasma (for detailed descriptions, 
see Hutchinson's review  \cite{hutchinson2024kinetic}). Valentini  \cite{valentini2012undamped} 
constructed a plateau distribution function with a flat region near 
the phase velocity to better describe the generation conditions of EAWs 
and used the Vlasov dispersion relation to identify the excitation region of undamped EAWs.

To explore whether the asymptotic state of collisionless 
plasma is a damping-free BGK mode, numerous theoretical  \cite{isichenko1997nonlinear,brodin1997nonlinear,lancellotti1998critical,lancellotti1999time}
and  simulation studies\cite{ghizzo1988stability,manfredi1997long,brunetti2000asymptotic,ivanov2004wave} on finite-amplitude perturbations  
have been carried out. These attempts have led to the conclusion 
that when a perturbation is applied to the distribution function 
and exceeds a threshold dependent on the initial conditions, the 
asymptotic state is a BGK mode formed by the superposition of two 
counter-propagating waves. This viewpoint has been mainstream for about 
20 years. However, recent advancements in computation  have 
enabled higher-precision, 
long-duration simulations that reveal more intricate structures.  \cite{carril2023formation,celebre2023phase}
In particular, the perturbation first decays 
as linear  damping, then forms a large vortex 
structure near the phase velocity in phase space. As time evolves, more small 
vortices form in regions below the phase velocity. These small 
vortices continue to evolve, and their energy spectrum follows 
 a power-law distribution similar to the Kolmogorov law.  \cite{nastac2025universal}

 In this work, we further refine Valentini's  \cite{valentini2012undamped}
analysis of the plateau distribution function, using the complete 
dispersion relation (we use the term dispersion relation even for Landau damped modes although they are not true eigenmode \cite{ng2004complete})
 in the complex plane. The perturbation of the electric field corresponding to 
the plateau distribution function is shown to form a weakly damped structure 
composed of two superimposed waves and the beat frequency is proportional 
to the size of the plateau in quasi-linear state. Using a second-order symplectic integrator, 
we perform high-resolution, long-time simulations of a finite-amplitude perturbation on a zeroth-order Maxwellian distribution. 
We establish that for this common class of perturbations, the plasma's asymptotic state is a stable, multi-wave BGK mode. 
The emergence of some of these BGK structures is found to be analogous to the generation of weakly damped waves in quasilinear theory.

The structure of this paper is as follows: Section \ref{section:2} presents the dispersion relation 
of the plateau distribution function in the complex plane and provides a 
comparison between the cases where the phase velocity of the velocity plateau coincides with 
and deviates from that corresponding to the Vlasov dispersion relation.  Section \ref{section:3} mainly compares the evolution 
of small-amplitude perturbations of the plateau distribution function with 
the evolution of Maxwellian distribution functions for
finite-amplitude perturbations. Section \ref{section:4} focuses on the multi-wave BGK 
structures with finite-amplitude perturbations. Section \ref{section:5} summarizes 
the paper and proposes possible applications for more complex systems.

\section{Plateau Distribution Function and Its Dispersion Relation} \label{section:2}

Based on the quasi-linear theory of weak turbulence, the normalized 
one-dimensional electrostatic plasma quasi-linear equations can be 
written as  \cite{Swanson2003quasilinear}

\begin{align}
  \frac{\partial f_1 }{\partial t }+v&\frac{\partial f_1 }{\partial x} - \frac{e}{m} E_1 \dfrac{\partial f_0}{\partial v} = 0 
  \label{equ1}
  \\\frac{\partial f_0}{\partial t} &= \frac{\partial}{\partial v} \left( D \frac{\partial f_0}{\partial v} \right)
  \label{equ2}
  \\ \frac{\partial E_1 }{\partial x }&=-\int^{\infty}_{-\infty}f_1 \mathrm{d}v \label{eq3}
  \\ E_{1,k}= &\frac{1}{2\pi}\int_{-\infty}^{\infty} E_1 e^{-ikx} \mathrm{d}x
  \\ D = &i \int_{-\infty}^{\infty}\frac{\tilde{E}_{1,k} \tilde{E}_{1,-k}}{\omega-kv}\mathrm{d}k
\end{align}
The time  normalization unit is the inverse of the plasma frequency \(\omega_p=\sqrt{4\pi n_0 e^2/m_e}\), the spatial normalization unit is the Debye length \(\lambda_d=\sqrt{T /4\pi n_0 e^2}\),
the velocity normalization unit is the thermal velocity\(\sqrt{T /m_e}\),
and the distribution function is normalized by the equilibrium particle 
number density. It is assumed that the ions are at rest and only the 
electron distribution function changes.

\begin{figure}[h]
    \centering 
    \includegraphics[width=\textwidth]{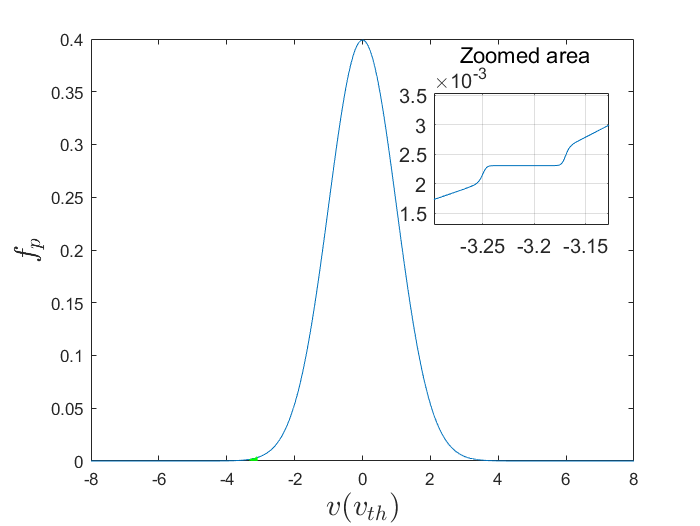}
    \caption{The plateau distribution function and its plateau area (zoomed area).}
    \label{fig:main1}
\end{figure}

From the above equations, it can be seen that the zeroth-order distribution function changes 
according to a diffusion equation, and the diffusion coefficient is 
closely related to the strength of the electric field perturbation.
Calculations  \cite{Swanson2003quasilinear} show that the diffusion term acts 
most strongly near the phase velocity. Based on quasi-linear theory, 
a plateau will be formed near the phase velocity in the asymptotic state, 
and its phenomenological expression can be written as  \cite{valentini2012undamped}
\begin{equation}
  \frac{f_p(v )}{N} =  f_M(v) - \frac{f_M(v) - f_M(v_0)}{1 + \left[ (v - v_0) / \Delta v_p \right]^{n_p}}- \frac{f_M(v) - f_M(v_0)}{1 + \left[ (v + v_0) / \Delta v_p \right]^{n_p}} 
\end{equation}
where \(f_M = \exp(-v^2/2)/\sqrt{2\pi}\) is the Maxwellian distribution, 
\(n_p\) is an even number, \(\Delta v_p = 0.02\), and \(N\) is a normalization constant 
slightly different from 1. It is important to note that \(f_p\) is smooth in \(v\),
and its derivatives up to the \(n_p \)-thorder vanish at \(v=v_0\). If 
\(n_p\) is sufficiently large, the distribution function can be 
approximated as flat within the interval \(\left[\pm v_0- \Delta v_p, \pm v_0 +\Delta v_p \right]\), forming a plateau distribution function (as shown in Figure \ref{fig:main1})

By combining equation (\ref{equ1}) and the Poisson equation, and 
performing analytical continuation with \(p=\omega_R+i\omega_I\) 
, the complete dispersion relation in the complex plane with \(\omega_I < 0\) consistent with linear theory can be obtained:
\begin{equation}
  k^2 -  \left[ \int_{-\infty}^{\infty} \frac{\partial f_p(v) / \partial v}{v - p/k} \, dv + 2i\pi \frac{\partial f_p}{\partial v} (p/k) \right] = 0
\end{equation}

The analytic continuation of \(f_p\) to the complex plane is natural, simply by 
replacing the real variable with a complex variable. Since this 
analytic continuation does not change the actual evolution of the electric field,  \cite{stucchi2025landau}
only this method of analytic continuation  is considered in this paper.

To accurately solve the dispersion relation, numerical calculations are required. 
Define the dispersion function: 
\begin{equation}
  \epsilon(p,k)= k^2 -  
\left[ \int_{-\infty}^{\infty} \frac{\partial f_p(v) / \partial v}{v - p/k} \, dv + 2i\pi \frac{\partial f_p}{\partial v} (p/k) \right] 
\end{equation}
In this work, the third-order Simpson integration formula is used, 
and an open formula is applied near the singularity for calculation.  \cite{press1992numerical_ch4}
The upper limit of the numerical integration is set at \(V_{max}=8\), and to 
precisely capture the effects of grid changes, integration uses \(N=4096\) data points, 
resulting in \(\Delta v=v_{max}/N\approx0.0039<\Delta v_p\). 

Similar to Valentini's  \cite{valentini2012undamped} work, a classification discussion is needed based on whether the
 phase velocity of the Vlasov dispersion relation is within the plateau.

\begin{figure}[h]
  \centering
   \captionsetup[subfigure]{labelformat=empty}
    \begin{subfigure}[t]{0.8\textwidth}
        \centering
        \includegraphics[width=\linewidth]{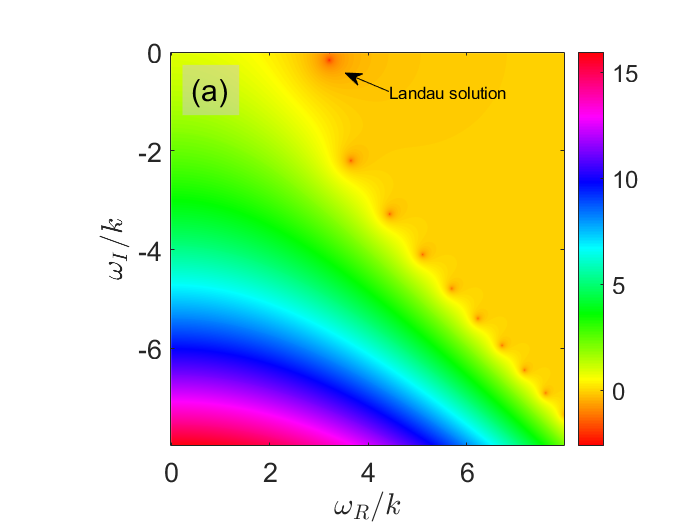}
        \caption{}
        \label{fig:2a}
    \end{subfigure}\\
    \begin{subfigure}[t]{0.8\textwidth}
        \centering
        \includegraphics[width=\linewidth]{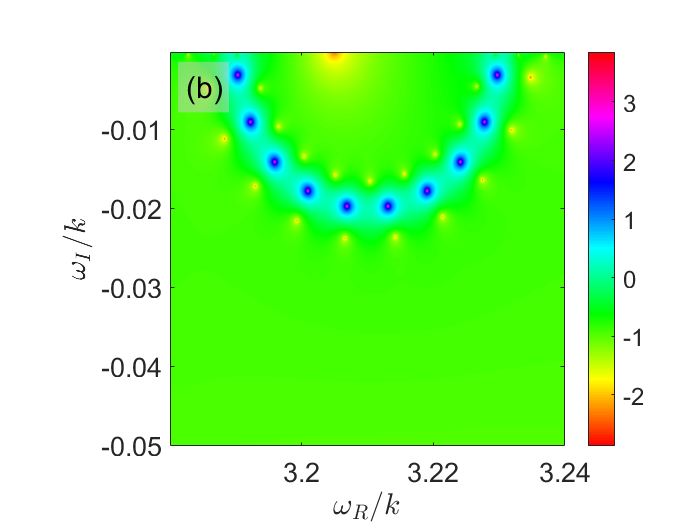}
        \caption{}
        \label{fig:2b}
    \end{subfigure}
  \captionsetup{justification=raggedright,singlelinecheck=false}
  \caption{ Base-10 logarithmic contours  of the 
  dispersion function for a plateau distribution with parameters 
  $k=0.4$,$v_0=3.21$ and $\Delta v_p=0.02,n_p=20$. (a) is the global dispersion function 
  and (b) is the dispersion function near the plateau.}
  \label{fig:main2}
\end{figure}

First, consider the case where the phase velocity is within the plateau.
Select \(v_0=3.21\) and \(k=0.4\), matching the phase velocity of the Vlasov dispersion 
relation, and set a small plateau  \(\Delta v_p=0.02\).
Since the plateau width is much smaller than the overall scale of the 
distribution function, it can be treated as a perturbation of the 
Maxwellian distribution. Figure \ref{fig:2a} reproduces the zero distribution of 
the dispersion relation corresponding 
to the Maxwellian distribution, where the uppermost zero point: \(\omega_R=1.284\), \(\omega_I=-0.640\) and \(v_p=\omega_R/k=3.21\)
is highly consistent with the Landau solution calculated from the linear theory dispersion function.  \cite{carril2023formation}
Moreover, a more intricate structure is found near the phase velocity plateau close to the real axis.
Figure \ref{fig:2b} is a zoomed area of the dispersion relation near the velocity plateau, showing a semi-circular structure 
with zeros near it.
The two lowest damped solutions are almost located at the edges of the plateau, suggesting that multi-wave superposition may 
form wave packets in the actual electric field evolution. Furthermore, a 
solution emerged  close to the real axis, indicating that in addition to 
the two weakly damped solutions, there exists an undamped solution.

\begin{figure}[htbp]
  \centering
  \captionsetup[subfigure]{labelformat=empty}
    \begin{subfigure}[t]{0.8\textwidth}
        \centering
        \captionsetup[subfigure]{labelformat=empty} 
        \includegraphics[width=\linewidth]{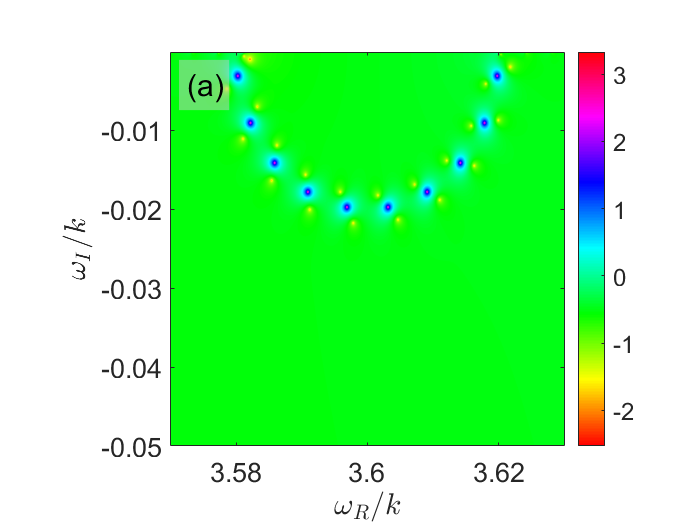}
        \caption{}
        \label{fig:3a}
    \end{subfigure}
    \\
    \begin{subfigure}[t]{0.8\textwidth}
        \centering
        \includegraphics[width=\linewidth]{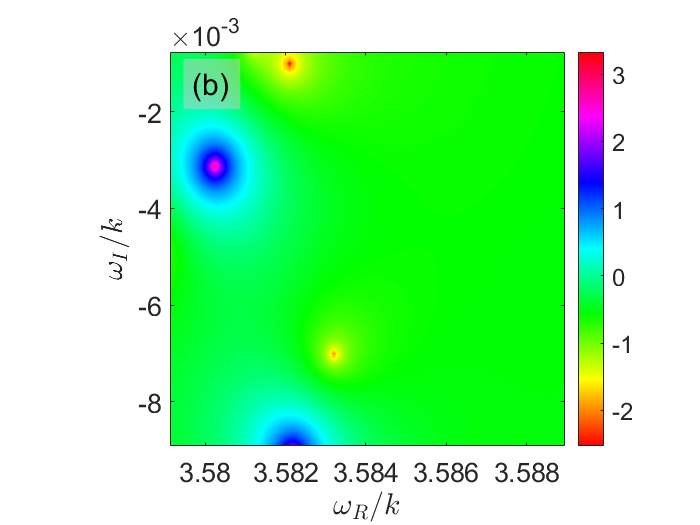}
        \caption{}
        \label{fig:3b}
    \end{subfigure}
  \captionsetup{justification=raggedright,singlelinecheck=false}
  \caption{ Logarithmic contours  of the dispersion 
  function corresponding to the plateau distribution function with 
  \(k=0.4,v_0=3.60\) and \(\Delta v_p=0.02\), \( n_p=20\). (a) is the dispersion function 
  near the plateau and
  (b) is near the weakest damping solution.}
  \label{fig:main3}
\end{figure}

Moving the velocity plateau to \(v_0=3.60\) while keeping other
parameters unchanged yields the new dispersion function near the 
plateau (as shown in Figure \ref{fig:main3}):

Compared with the dispersion function when the phase velocity is 
within the plateau, when the phase velocity does not coincide with the 
velocity plateau, only the solutions near the semi-circular structure 
remain (Fig. \ref{fig:3a}), and the solution near the real axis disappears, consistent with 
Valentini's  \cite{valentini2012undamped} conclusion considering only the Vlasov dispersion relation.

It can be reasonably inferred that after the system undergoes the initial Landau damping and reaches
 the asymptotic state, a series of weakly damped solutions will remain. When the phase velocity 
coincides with the plateau, two lowest-damped solutions and \added{an} undamped solution will remain: 
the nearly damped-free solution near the real axis and two pairs of 
weakly damped solutions at the edges of the plateau. When the phase 
velocity does not coincide with the plateau,
 only two  weakly damped solutions will remain.

 \section{Comparison Between  Evolutions of Small-Amplitude Perturbation to  Plateau Distribution
  and Finite-Amplitude Perturbation to Maxwellian Distribution} \label{section:3}

\begin{figure}[h] 
    \centering 
    \captionsetup[subfigure]{labelformat=empty}
    \begin{subfigure}[t]{0.8\textwidth} 
        \centering
        \includegraphics[width=\linewidth]{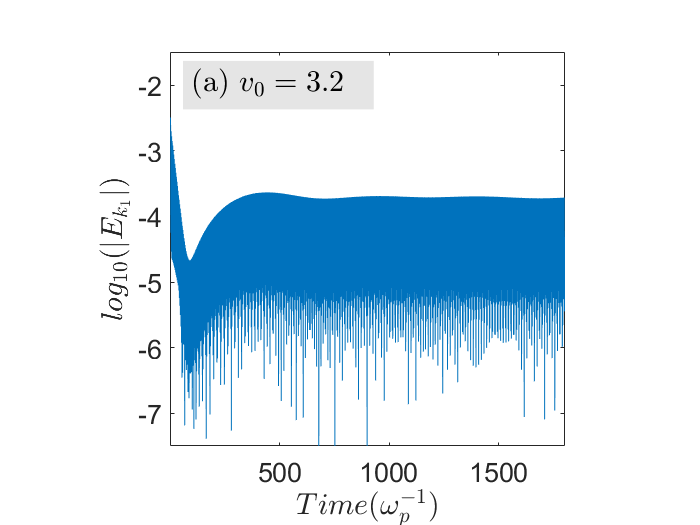} 
        \caption{}
        \label{fig:4a}
    \end{subfigure}\\
    \begin{subfigure}[t]{0.8\textwidth}
        \centering
        \includegraphics[width=\linewidth]{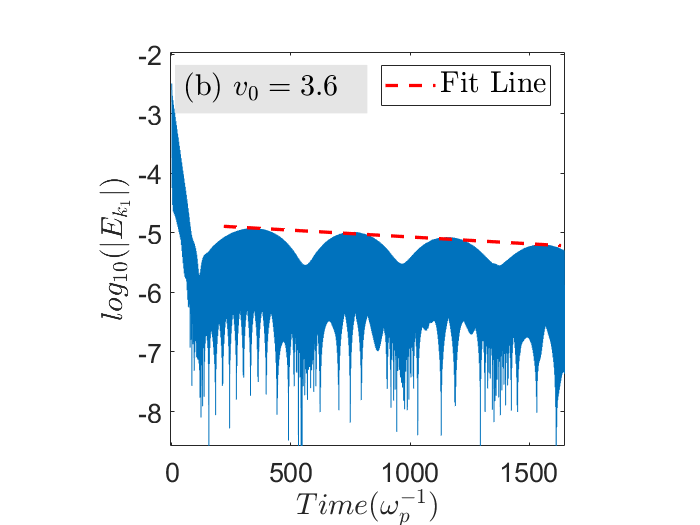} 
        \caption{}
        \label{fig:4b}
    \end{subfigure}
    \captionsetup{justification=raggedright,singlelinecheck=false}
    \caption{ The evolution of the electric field energy with small-amplitude initial perturbations, 
     where the phase velocity of the velocity plateau (a) does and (b) does not  coincide with that of the Vlasov dispersion relation, while  (b) shows the case where it does not.}
    \label{fig:main4}
\end{figure}

 The one-dimensional Vlasov-Poisson system can be numerically solved on the phase plane
\((x,v)\) using a second-order symplectic  algorithm with cubic spline interpolation,  \cite{cheng1976integration,watanabe2005vlasov}
which preserves the symmetry of the phase space distribution function under symmetric perturbation, i.e., \(f(x,v,t)=f(L-x,-v,t)\)
 (see Appendix \ref{appendixb} for details).
Specifically, the phase space is discretized onto a uniform Cartesian grid over the rectangular region 
\(R\equiv \{ (x, v) | 0 \leq x < L_x, |v| \leq v_{\text{max}} \}\),
 where \(L_x\) is the periodic spatial length and \(v_{max}\) is the  
 cutoff velocity. The space coordinate \(x\) and velocity \(v \) are discretized into uniform
 grid sizes \(\delta x =L_x/ N_x \) and \(\delta v=2v_{max}/N_v \), 
 with \(N_x \) and \(N_v\) being the number of grid cells in the \(x \) and \(v \) directions respectively.
 In Fourier space, the wavenumber is defined as \(k_n=2\pi n/L_x \), where \(n=0,\ldots,N_x /2 \).
 For the plateau distribution function, the initial condition including perturbation can be expressed as

\begin{equation}
F_p(x, v, t=0) = f_p+\alpha f_M \cos(k_1x)
\label{equ5}
\end{equation}
where \(\alpha=10^{-5}\) is used to determine the initial perturbation  strength, and
\(k_1=0.4\) is the initial perturbation wavenumber, corresponding to the fundamental
Fourier mode in the wavenumber domain (i.e., the largest scale). To achieve sufficient accuracy, we choose \(N_x=256\) and \(N_v=4096\).
The plateau distribution function uses the same parameters as in the previous section for calculating the dispersion relation. The evolution of the electric field with 
small-amplitude perturbation of the plateau distribution function is shown in Fig.\ref{fig:main4}

We start with a resonant case where the plateau location matches the wave phase velocity.
From Fig. \ref{fig:4a}, it can be seen that when a plateau exists, the electric field evolution follows the linear Landau damping theory 
in the early stage of the simulation. After about 100 time units, the amplitude begins to rebound and eventually 
reaches a steady state. Careful observation of the electric field evolution in the steady state reveals that the electric field amplitude evolution 
in the steady state forms a wave packet structure with a period of about 488 time units.

From Fig. \ref{fig:4b}, it can be seen that when the plateau location does not match the phase velocity, the electric field evolution also follows the linear Landau damping theory in the early stage. 
However, after reaching the steady state, the electric field evolution is not a stable wave packet but a decaying wave packet. Fitting can yield a damping rate   \(\propto\exp(-4.6\times 10^{-4}\omega_p t)\),
which is consistent with the damping rate  \(\omega_I=-4\times 10^{-4}\) corresponding to the weakest damping solution in Fig. \ref{fig:3b}. Since the dispersion relation has multiple weakly damped solutions, 
the slight differences between the simulation and the numerical dispersion relation may originate from the influence of  strong damping solutions.

According to Appendix \ref{appendixa}, in the limit of \(n_p\gg 1\), 
an analytical solution of the dispersion relation can be obtained. In the zeroth-order approximation, the period of the wave packet can be derived as 
\(T=\pi/(2\Delta v_p\cdot k)\) (The factor of \(1/2\)
 is because Figure \ref{fig:main4} shows the absolute value of the Fourier component with \(k_1=0.4\)).
Substituting the data yields \(T=392.6\) time units, while the period calculated from the numerical dispersion relation is 473.1 time units, 
which are relatively close. This further verifies the accuracy of the numerical calculation results.

\begin{figure}[h]
    \centering 
    \includegraphics[width=\textwidth]{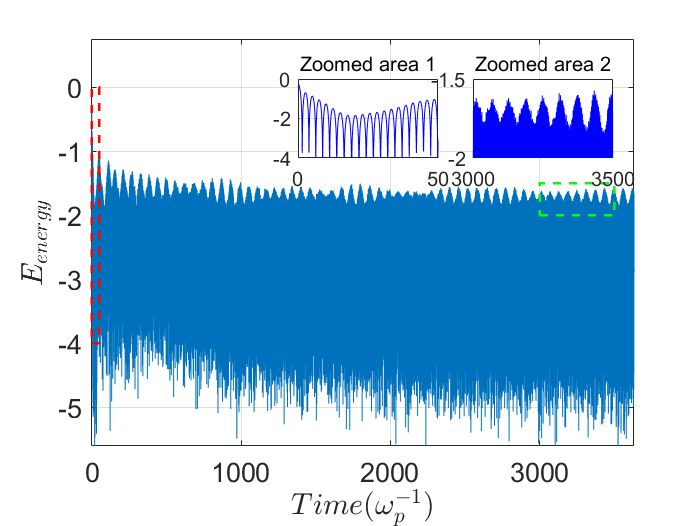}
    \captionsetup{justification=raggedright,singlelinecheck=false}
    \caption{ Long-time evolution of the electric field energy with finite-amplitude initial perturbation with parameters \(\beta=0.15 ,k_1=0.4\).}
    \label{fig:main5}
\end{figure}

For the evolution of a finite-amplitude perturbation to the Maxwellian distribution, a similar initial perturbation is adopted:

\begin{equation}
  F_M(x, v, t=0) = f_M\left[1+\beta \cos(k_1x)\right]
\end{equation}
where \(\beta=0.15\) is used to determine the initial perturbation strength, and 
\(f_M \) is the Maxwellian distribution function. To capture the fine structures,
 \cite{carril2023formation} we choose \(N_x=2048\) and \(N_v=16384\). 
In the steady state, a more complex wave packet structure different from the double-wave BGK structure
 \cite{ghizzo1988stability,manfredi1997long,brunetti2000asymptotic,ivanov2004wave} is also observed:

Figure \ref{fig:main5} shows the long-time evolution of the electric field energy with  \(\beta=0.15 ,k_1=0.4\).
The physical quantity on the vertical axis, \(E_{energy}=\frac{1}{2}\int\left|E\right|^2 dx\), 
represents the  electric field energy in a period domain. 
In the early stage of evolution (see the zoomed area 1 in Fig.\ref{fig:main5}),
the electric field exhibits an initial exponential decay at a 
rate approximately 1.5 times that of linear Landau damping, an 
enhancement attributed to nonlinear mode coupling (energy transfer to higher harmonics).\cite{cheng1976integration,o1965collisionless} 
Subsequently, driven by trapped particle, the field begins to grow and eventually reaches a steady state. (see the zoomed area 2 in Fig.\ref{fig:main5}). The steady-state electric field energy evolution can roughly be seen to consist of two  scales: one 
is the rapid oscillation close to the Langmuir frequency, and the other is the envelope of the wave packet structure. This is similar to the steady-state 
of the plateau distribution function under small-amplitude evolution in Sec \ref{section:2}. However, as shown in the analysis in the next section, 
the steady-state of finite amplitude has a more complex 
frequency spectrum distribution.

\section{Multi-Wave BGK Modes in the Steady State of Collisionless Plasma}\label{section:4}
\begin{figure*}[ht]
  \centering
  \captionsetup[subfigure]{labelformat=empty}
  \begin{subfigure}{0.5\textwidth}
    \centering
    \includegraphics[width=\linewidth]{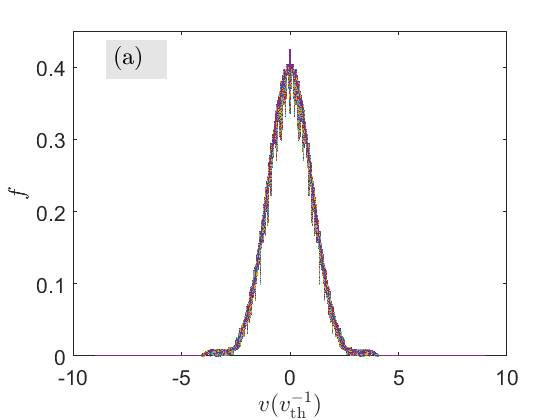}
    \caption{}
    \label{fig:6a}
  \end{subfigure}\\
  \begin{subfigure}{0.5\textwidth}
    \centering
    \includegraphics[width=\linewidth]{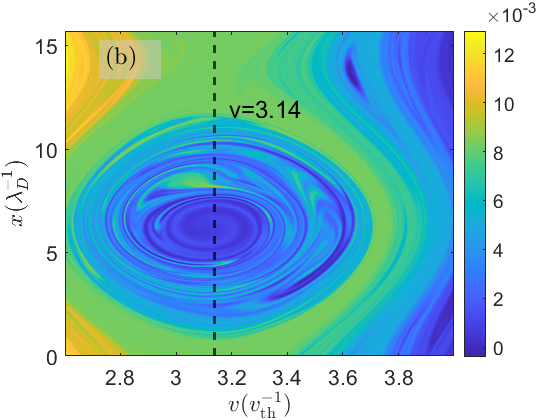}
    \caption{}
    \label{fig:6b}
  \end{subfigure}\\
  \begin{subfigure}{0.5\textwidth}
    \centering
    \includegraphics[width=\linewidth]{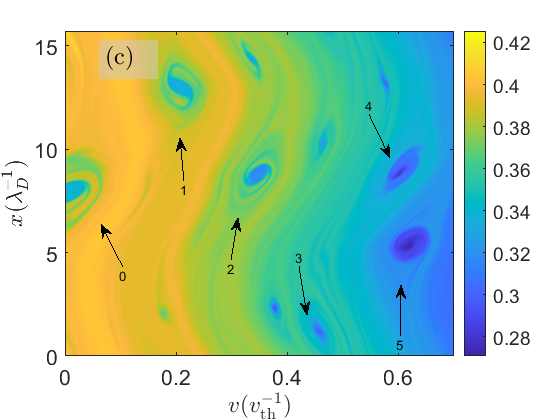}
    \caption{ }
    \label{fig:6c}
  \end{subfigure}\\
  \captionsetup{justification=raggedright,singlelinecheck=false}
  \caption{Phase space distribution function of the asymptotic state with \(\beta=0.15,k=0.4\),\(t=3750\omega_p^{-1}\).
  (a) is a projection diagram of distribution function along the x-direction, (b) presents zeroth-order vortex in phase space and
   (c) presents small vortices near zero velocity in phase space.}
  \label{fig:main6}
\end{figure*}

For the asymptotic state of a one-dimensional electrostatic plasma, consider the phase space structure of the distribution function 
at \(\beta=0.15\), \(k=0.4\), \(t=3750\omega_p^{-1}\)
(Figure \ref{fig:main6}). The projection of the steady-state distribution 
function in the \(x\) direction (Fig. \ref{fig:6a}) mainly undergoes strong 
distortions in two regions compared to the Maxwellian distribution function. One is the result of 
the  Langmuir wave, which forms a broad plateau region at the corresponding phase velocity (The presence of nonlinear chirping causes the phase velocity of the vortex center to shift down from 3.21 to 3.14\cite{morales1972nonlinear,berger2013electron}). However, unlike the quasi-linear theory  \cite{Swanson2003quasilinear}
and the assumption of the plateau distribution function in the previous section, this plateau region in 1D projection has a more complex structure and corresponds to a large vortex structure in 
 2D projection phase space (Fig.\ref{fig:6b});
the other is the region near zero phase velocity, where there are multiple narrow and deep vortices, as shown in Fig. \ref{fig:6c}. Due to the effect of the zeroth-order vortex, 
an open wave-like structure is formed outside the vortex (similar to the phase space structure corresponding to a sinusoidal potential), 
and these small vortices are embedded in these open wave-like structures and act as perturbations without affecting the topological structure of most regions of phase space.

Although the asymptotic state is characterized by these complex vortex structures rather than a simple plateau, 
we still observe spectral features in the electric field that are analogous to those predicted for a plateau distribution. 
As shown in Fig.\ref{fig:6a}, for the electric field component with wavenumber \(k=0.4\), 
two nearly symmetric sidebands are present on either side of the main zeroth-order frequency component at \(Frequency =1/T \approx 0.2\omega_p\). 
These sidebands act as a first-order perturbation, and their beating is the source of the envelope modulation observed in the asymptotic electric field energy 
in Fig.\ref{fig:main5} of Sec \ref{section:3}.

\begin{figure}[b]
   \centering
    \includegraphics[width=\textwidth]{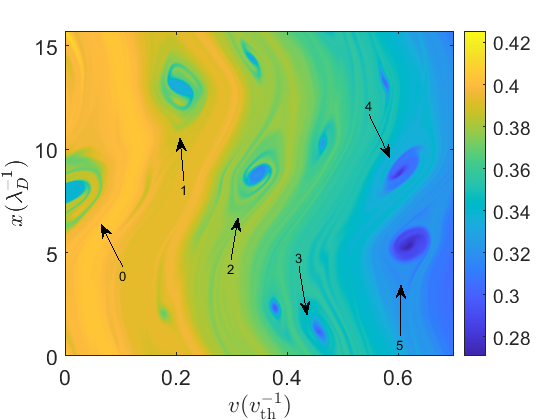}
    \captionsetup{justification=raggedright,singlelinecheck=false}
    \caption{Vortex structure of the distribution function in phase space. 
    The arrows indicate the locations corresponding to the discrete spectral peaks \(\omega_i\) shown in Fig. 7.}
    \label{fig:main8}
  \end{figure}

Furthermore, a series of discrete peaks is evident in the low-frequency region (see zoomed area in Fig.\ref{fig:main7}). These peaks correspond to 
the vortices located in the low-velocity region of phase space. To demonstrate this connection, we convert the frequencies  of several prominent peaks 
into phase velocities via $v_{pk} = \omega/k$ and mark the corresponding locations in phase space with arrows in Fig.\ref{fig:main8}. 
The markers align well with the vortex.

According to the theory of superposed multi-wave BGK modes,\cite{buchanan1993superposition} in the weakly nonlinear, zeroth-order approximation, the system's state can be modeled as a superposition of independent BGK modes. 
A defining characteristic of a single-wave BGK mode is its well-defined phase velocity. 
This theoretical framework predicts that the discrete spectral lines \(\omega_i\) observed at \(k_E = 0.4\) (for \(i=0, \dots, 4\)) 
should have counterparts in the spectra of higher wavenumbers, such as \(k_E = 0.8\) and \(k_E = 1.2\). 
These counterparts should appear at frequencies that maintain the same phase velocity, i.e., $v_{p,i} = \omega_i/k_E$ is equivalent for each mode. 
Indeed, we observe corresponding spectral lines at higher wavenumbers.
Moreover, the spectral amplitudes for these modes in the low-phase-velocity region are of the same order of magnitude across all three wavenumbers. 
This observation contradicts the predictions of weakly nonlinear multi-wave interaction theory,\cite{Swanson2003finiteamplitude} which would anticipate an  order-of-magnitude separation between 
the spectral amplitudes for these different wavevectors. 
This discrepancy suggests that even the zeroth-order BGK theory, which neglects  interactions between BGK structures, cannot adequately explain 
the origin of these vortices by weakly nonlinear multi-wave interaction theory.

To quantitatively assess the results, we use the relation \(v_{pk} = 2\pi f/k_E\)
 to convert the spectral peak frequencies (\(\omega\)) into theoretical phase velocities (\(v_{pk}\)). 
 We compare these with the actual vortex velocities, which we define as the velocity at which the projected distribution function reaches a local minimum \
 (the depth of the vortices relative to their width, as seen in Fig.\ref{fig:6a}, validates this assumption). 
 The comparison is summarized in Table \ref{table1}. We find a  small discrepancy between the predicted phase velocities 
 and the actual vortex locations. This indicates that while the single-wave BGK approximation is reasonably accurate for the low-velocity structures, 
 the true asymptotic state is a fully self-consistent system involving the entire phase-space structure and the total electric field.

However, for the sideband spectra flanking the main frequency, we do not observe distinct vortex structures. 
Instead, the phase-space region near the Langmuir wave phase velocity (Fig.\ref{fig:6b}) is occupied by a single, large vortex with a complex internal structure. 
We hypothesize that strong nonlinear effects have caused the  vortices associated with the sidebands and the main Langmuir wave to merge, 
forming a new, unified, and self-consistent structure that is not a simple single-wave BGK mode. 

\begin{figure}[htb]
    \centering 
    \captionsetup[subfigure]{labelformat=empty}
    \begin{subfigure}[b]{0.23\textwidth}
        \includegraphics[width=\textwidth]{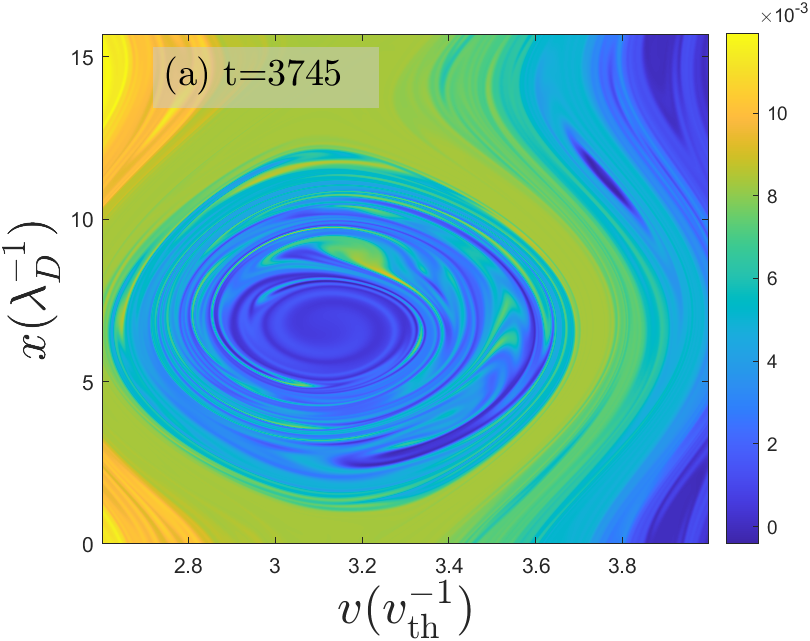}
        \caption{}
        \label{fig:9a}
    \end{subfigure}
    \begin{subfigure}[b]{0.23\textwidth}
        \includegraphics[width=\textwidth]{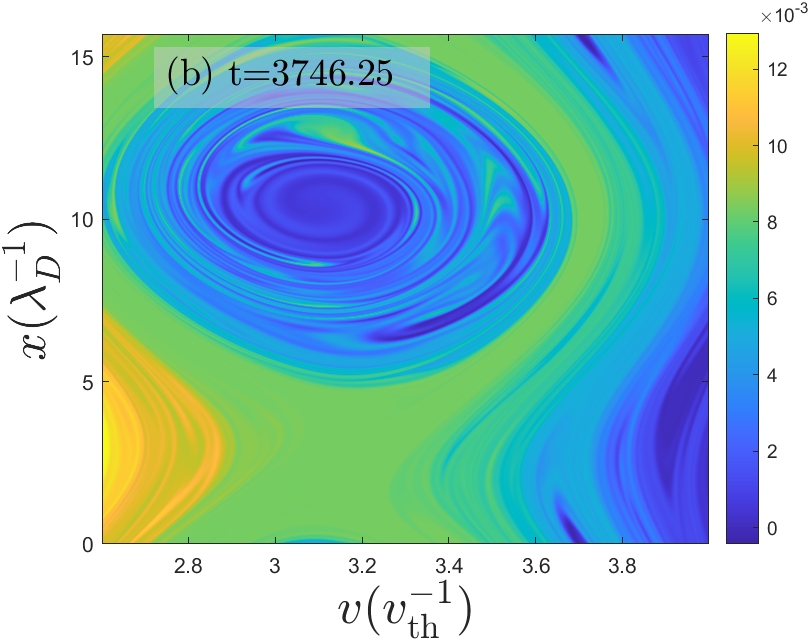}
        \caption{}
        \label{fig:9b}
    \end{subfigure}
    \begin{subfigure}[b]{0.23\textwidth}
        \includegraphics[width=\textwidth]{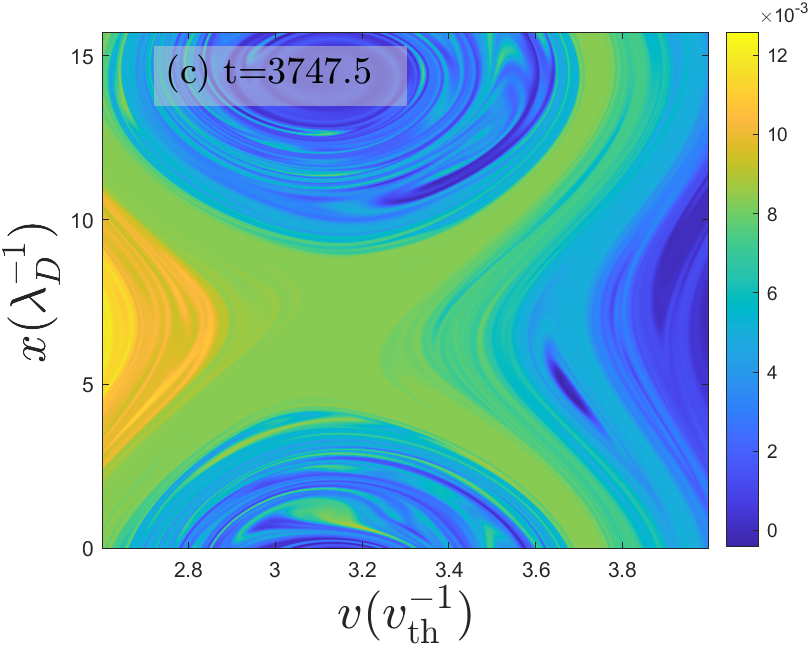}
        \caption{}
        \label{fig:9c}
    \end{subfigure}
    \begin{subfigure}[b]{0.23\textwidth}
        \includegraphics[width=\textwidth]{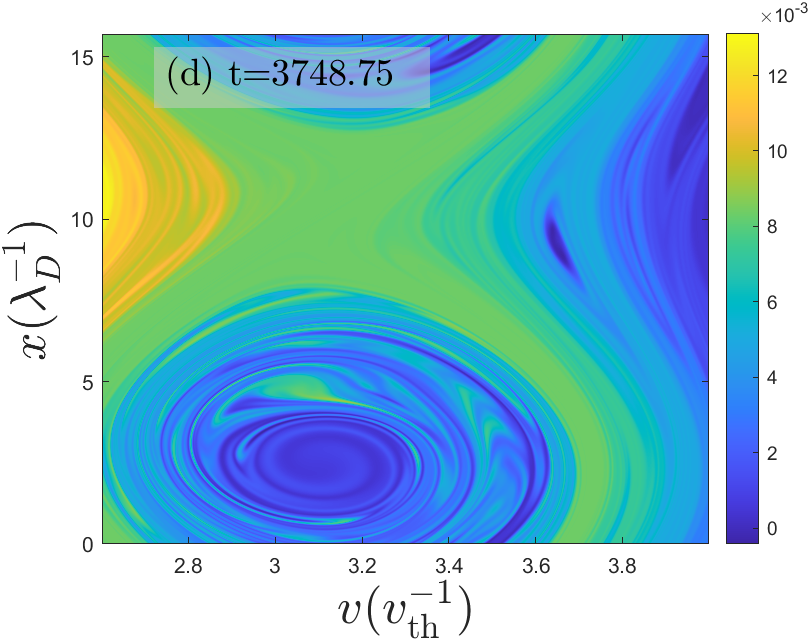}
        \caption{}
        \label{fig:9d}
    \end{subfigure}

    \vspace{1cm} 
    \begin{subfigure}[b]{0.3\textwidth}
        \includegraphics[width=\textwidth]{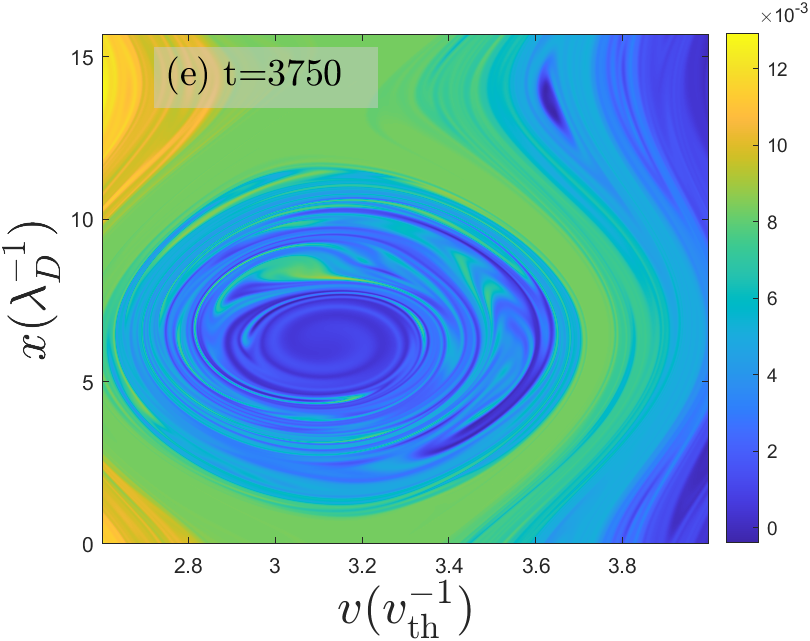}
        \caption{}
        \label{fig:9e}
    \end{subfigure}
    \begin{subfigure}[b]{0.3\textwidth}
        \includegraphics[width=\textwidth]{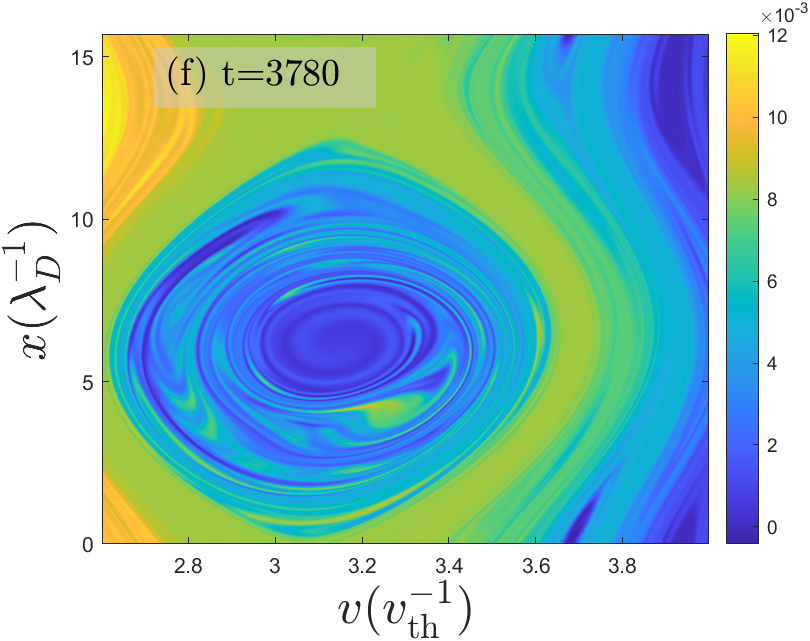}
        \caption{}
        \label{fig:9f}
    \end{subfigure}
    \begin{subfigure}[b]{0.3\textwidth}
        \includegraphics[width=\textwidth]{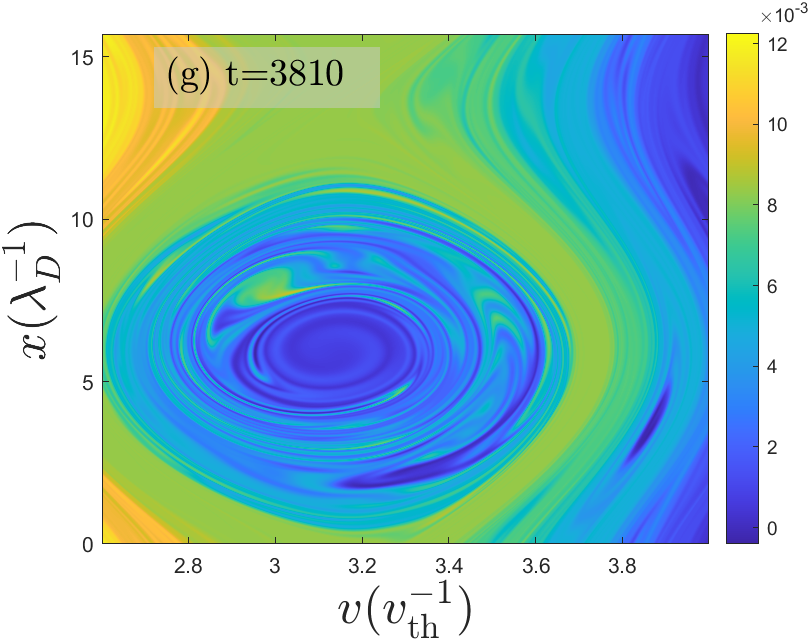}
        \caption{}
        \label{fig:9g}
    \end{subfigure}

    \caption{Snapshots of the distribution function contours illustrating the internal rotation of the large vortex. Panels (a)-(e) show the evolution over one transit period. 
    The comparison between (e) and (g) shows the near-recurrence of the structure after 12 transit periods, corresponding to one envelope period.}
    \label{fig:main9}
\end{figure}

If this large vortex structure were a single-wave BGK mode, its period of motion in phase space, according to the theory of 
single-wave BGK mode \cite{bernstein1957exact}, would be given by \(T=L_x /v_{pm}=5 \omega_p^{-1}\), where  \(v_{pm}=3.14\) is the velocity 
of the large vortex's center. Figures \ref{fig:9a}-\ref{fig:9e} present a series of snapshots of the distribution function contours 
over one such period. It is observed that after one period, the vortex center returns to its initial position. However, 
its internal structure exhibits a net rotation by a certain angle relative to the center.

This internal rotation is responsible for the periodicity of the electric field energy envelope, 
which is observed in the second zoomed area of Fig. \ref{fig:main5} and has a period of approximately \( 61 \omega_p^{-1}\). 
To verify this connection, we compare the snapshot of the distribution function at \(t=3750\omega_p^{-1}\) (Fig. \ref{fig:9e}) with the one at \(t=3810\omega_p^{-1}\) (Fig. \ref{fig:9g}), 
a time that corresponds to the passage of 12 vortex transit periods (\(12\times T=60\omega_p^{-1}\)). It is evident that the vortex structures in the two snapshots are 
nearly identical. This demonstrates that the envelope period of the electric field energy is indeed identical to the rotational period of the large vortex.

Taken together, the overall structure can be effectively approximated as a superposition of a large, 
 complex vortex at the Langmuir phase velocity and several smaller, quasi-independent vortices at low velocities.

\begin{table*}
    \caption{Comparison between the theoretically predicted phase velocity and the actual vortex phase velocity}
     \label{table1}
     \begin{ruledtabular}
        \begin{tabular}{cccccccc}
            Peak Number &0& 1 & 2 & 3 & 4 &5 \\ \hline
            \(k_E=0.4\) &0& 0.167 & 0.328 & 0.475 & 0.579  &0.663\\ 
            \(k_E=0.8\) &0& 0.172 & 0.331 & 0.478 & 0.580  &0.660 \\
            \(k_E=1.2\) &0& 0.170 & 0.330 & 0.477 & 0.582  &0.663\\ 
            Actual Vortex &0& 0.193 & 0.345 & 0.457 & 0.571 &0.618 \\ 
        \end{tabular}
     \end{ruledtabular}
\end{table*}

\begin{figure*}[t]
  \centering
  \captionsetup[subfigure]{labelformat=empty}
  \begin{subfigure}{0.5\textwidth}
    \centering
    \includegraphics[width=\linewidth]{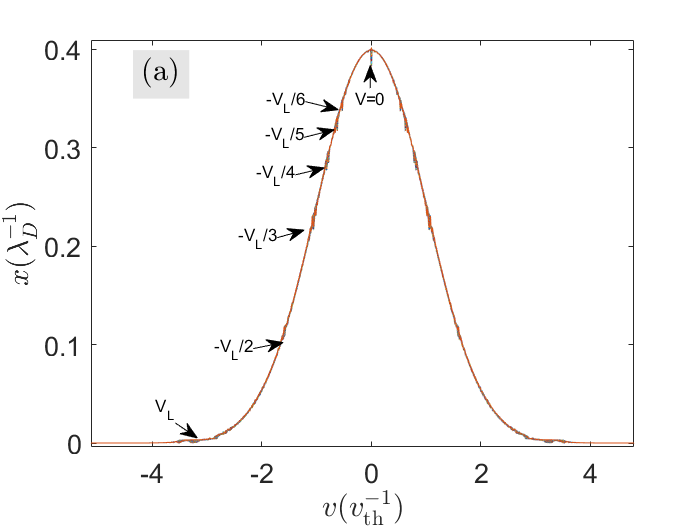}
    \caption{}
    \label{fig:10a}
  \end{subfigure}
  \\
  \begin{subfigure}{0.5\textwidth}
    \centering
    \includegraphics[width=\linewidth]{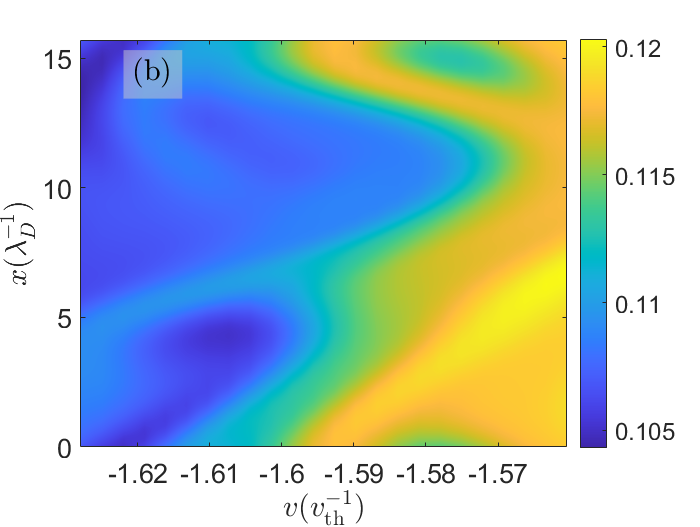}
    \caption{}
    \label{fig:10b}
  \end{subfigure}
  \\
  \begin{subfigure}{0.5\textwidth}
    \centering
    \includegraphics[width=\linewidth]{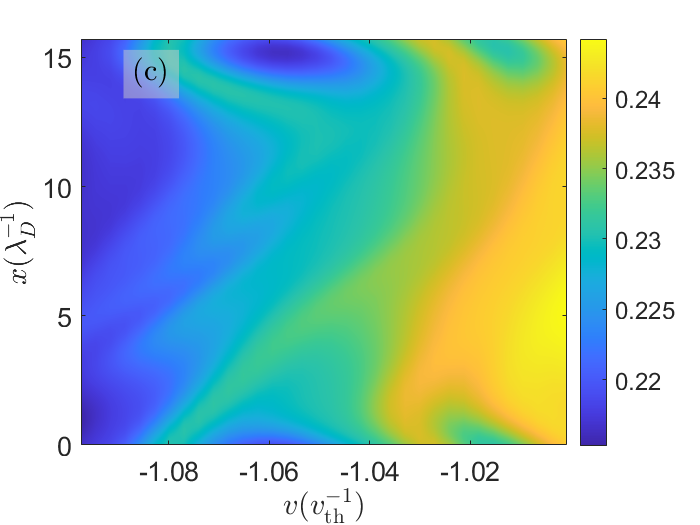}
    \caption{}
    \label{fig:10c}
  \end{subfigure} 
  \captionsetup{justification=raggedright,singlelinecheck=false}
  \caption{The distribution function corresponding to 
\(\beta=0.04,k=0.4\),\(t=2500\omega_p^{-1},N_x=128,N_v=4096\), a presents small vortex structures at \(v_p/n\),
(b) and (c) show the phase space vortex structures at \(v_p/2\) and \(v_p/3\), respectively.}
   \label{fig:main10}
\end{figure*}

 In addition to the phase space vortex generated by the nonlinear interaction of the multi-wave,
  Carril  \cite{carril2023formation} also discovered small vortex structures corresponding to \(v_p/n\).
   Similarly in simulations with \(\beta=0.04\),\(N_v=4096\), and low spatial resolution \(N_x=128\),
small vortex structures at \(v_p/n\) can also be observed (Figure \ref{fig:main10}).
From Figure \ref{fig:10a}, the small vortices at \(v_p/n\)
 can be clearly seen by  the projection of the distribution function onto the \(v\)-direction. 
 Figures \ref{fig:10b} and \ref{fig:10c} show the 
 phase space structures near \(v_p/2\) and \(v_p/3\).
 In Figure \ref{fig:10b}, the two vortices are strongly distorted due to strong coupling, making 
 the positions of the vortex centers difficult to determine. In contrast, 
 the positions of the vortex centers in Figure \ref{fig:10c} can be well determined. 

\begin{figure*}[t]
  \centering
  \captionsetup[subfigure]{labelformat=empty}
  \begin{subfigure}{0.375\textwidth}
    \centering
    \includegraphics[width=\linewidth]{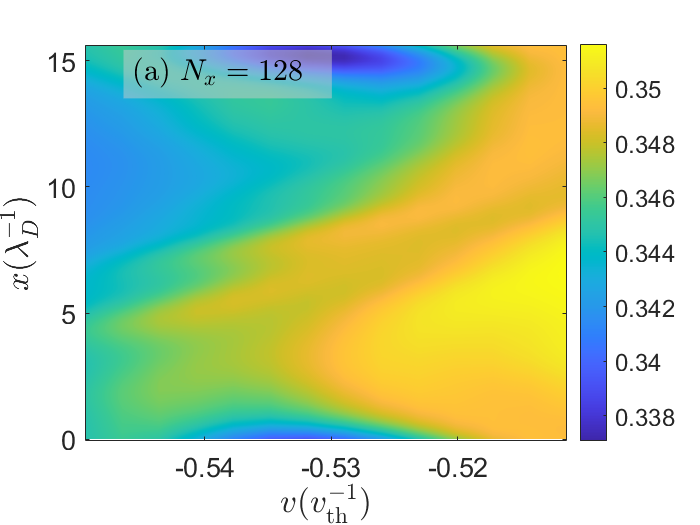}
    \caption{}
    \label{fig:11a}
  \end{subfigure}
  \\
  \begin{subfigure}{0.375\textwidth}
    \centering
    \includegraphics[width=\linewidth]{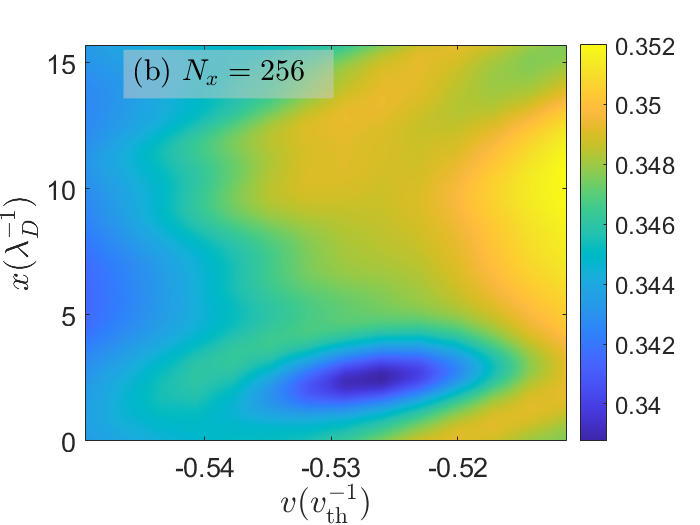}
    \caption{}
    \label{fig:11b}
  \end{subfigure}
  \\
  \begin{subfigure}{0.375\textwidth}
    \centering
    \includegraphics[width=\linewidth]{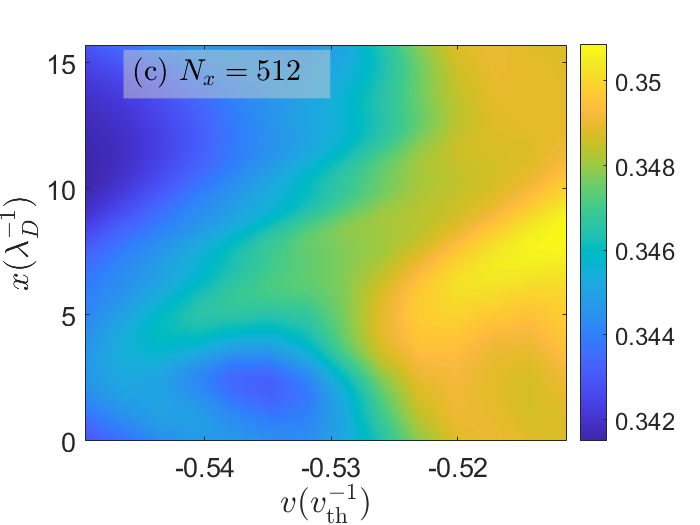}
    \caption{}
    \label{fig:11c}
  \end{subfigure}
  \\
  \begin{subfigure}{0.375\textwidth}
    \centering
    \includegraphics[width=\linewidth]{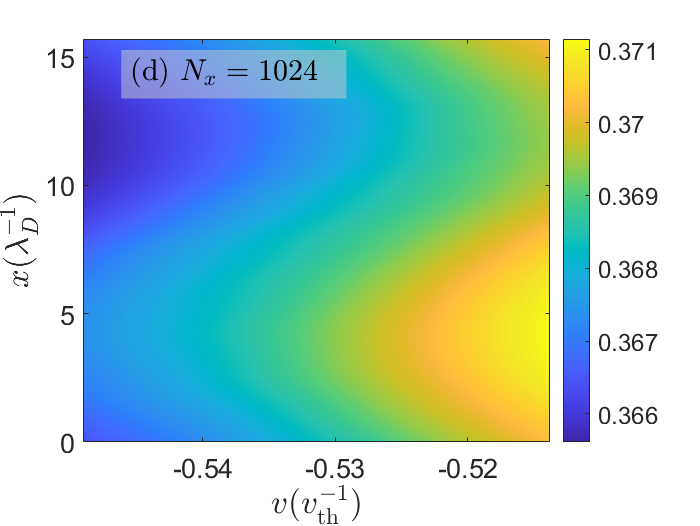}
    \caption{}
    \label{fig:11d}
  \end{subfigure}
  \captionsetup{justification=raggedright,singlelinecheck=false}
  \caption{For \(\beta=0.04,k=0.4\),\(t=2500\omega_p^{-1},N_v=4096\) the changes in phase - space vortices at
  \(V_L/6\) with the variation of \(N_x \), (a)-(d) present the process by which the vortex disappears as the \(N_x\) increases.}
  \label{fig:main11}
\end{figure*}

  However, as the spatial resolution increases, the phase 
 space vortices gradually decrease in size and eventually disappear 
 (Figure \ref{fig:main11}). Figure \ref{fig:main11} illustrates how the phase space vortices at \(v_p/6\) 
change as \(N_x \) increases. As the spatial resolution improves, the vortices at \(v_p/6\)
gradually shrink and eventually disappear. Similar behavior is observed at other \(v_p/n\)
positions, but the vortices in the low-phase-velocity region remain 
unaffected. This indicates that this phenomenon is numerical 
in nature, likely originating from numerical dispersion during the 
Fourier transform in the x direction. That is, single-wave perturbation 
generates higher wavenumber fluctuations through 
Fourier transformation. This acts as a driving force at \(v_p/n\)
phase velocities, naturally producing vortices at these velocities. 

For small-amplitude steady states, the vortices generated at 
low phase velocities are weak, making this numerical phenomenon more pronounced 
in the steady states of small-amplitude perturbations. In contrast, 
the impact of this phenomenon is less significant in the steady states 
of large-amplitude perturbations. 
It should also be noted that even in Carril's results \cite{carril2023formation}, 
with \(N_v \)  reaching 131,072 grids, this phenomenon was not avoided. This may be due to the spatial resolution
\(N_x =2048\) being nearly two orders of magnitude lower than \(N_v\), 
allowing even small numerical dispersion to generate fine structures in phase space. To avoid this phenomenon, it is necessary to reasonably match 
\(N_v\) and \(N_x\).

\section{Conclusion}\label{section:5}

This study combines analytical methods and numerical simulations to investigate the  state of nonlinear Landau damping in one-dimensional collisionless plasma.

In Sec \ref{section:2},  following the work of Valentini  \cite{valentini2012undamped}
on plateau distribution functions, we assume that the initial distribution function has a small but non-zero plateau. By analytically extending the dispersion relation 
into the complex plane, we discuss two scenarios: when the phase velocity of the Vlasov dispersion relation coincides with the velocity plateau and when it does not. 
When the phase velocity coincides with the velocity plateau, there exists a nearly undamped solution on the real axis and weakly damped solutions at the edges of the 
plateau. When the phase velocity does not coincide with the velocity plateau, the undamped solution on the real axis disappears, 
leaving only weakly damped solutions at the edges of the plateau.

In Sec \ref{section:3}, we compare the evolution of the near-zero amplitude perturbation applied to the plateau distribution (with \(\alpha=10^{-5}\))
with that of the finite-amplitude perturbation applied to the Maxwellian distribution  (with \(\beta=0.15\)).
In the asymptotic state, there exists a zeroth-order asymptotic electric field, which represents the rapid oscillation of Langmuir waves, and there is a more refined 
structure, namely the secondary asymptotic electric field, resulting in the wave packet structure of the envelope. Furthermore, for the near-zero amplitude evolution,
 the weakly damped solutions in the steady state align well with the solutions of the dispersion function analytically extended into the complex plane in Sec \ref{section:2}. 
We speculate that the multi-wave superposition in the finite-amplitude asymptotic state may arise from the multi-solutions of the dispersion relation caused by 
the distortion in the distribution function near the phase velocity.

Sec \ref{section:4} investigates the asymptotic state of nonlinear Landau damping, identifying it as a complex multi-wave Bernstein-Greene-Kruskal (BGK) structure. 
Our analysis reveals that this state is characterized by two distinct 
features in phase space: a large, complex vortex near the Langmuir wave 
phase velocity after nonlinear chirping, which has a rotational period  
synchronized with that of the electric field envelope; and a series of smaller, discrete vortices at low velocities.
Through spectral analysis of the electric field, we demonstrate a direct correspondence between 
these low-velocity vortices and a series of discrete, low-frequency peaks.

While the existence of modes with well-defined phase velocities is formally consistent with zeroth-order multi-wave BGK theory, the observed amplitude scaling 
across different wavenumbers contradicts the predictions of weakly nonlinear interaction theory, indicating the crucial role of strong nonlinear effects. 
Finally, this section clarifies that the \(v_L/n\) vortex structures reported in previous literature 
are numerical artifacts dependent on spatial resolution, distinguishing them from the physically robust low-velocity vortices identified in this work.

To sum up, this study provides a new understanding of the asymptotic state of nonlinear Landau damping in one-dimensional plasmas. 
The results highlight the importance of nonlinear interactions and the role of phase-space vortices in shaping the long-term behavior of the system. 
Future work should focus on: (1) providing a quantitative explanation for the formation mechanism of vortex structures in 
the low phase-velocity region and the rotational dynamics of the large vortex ;
(2) extending these findings to more complex systems, such as high-dimensional plasmas and magnetized plasmas,
to gain deeper insights into the fundamental processes governing nonlinear Landau damping.

\begin{acknowledgments}
We are grateful for the support from the  D.F.Escande. This work was supported
by the National Magnetic Confinement Fusion Program of
China (Grant No. 2019YFE03050004),  the Hubei International Science and Technology Cooperation Project under Grant No. 2022EHB003, and U.S.
Department of Energy (Grant No. DE-FG02-86ER53218).
The computing work in this paper is supported by the Public Service 
Platform of High Performance Computing by Network and Computing Center of HUST.
C. -S. Ng was supported by a U. S. National Science Foundation Grant PHY-2511536.
We would also like to express our gratitude to  Jiaxing Liu and Weimin Fu for their
 valuable help and discussions throughout this project.
\end{acknowledgments}

\appendix
\section{The Weakly Damped Solution of the Plateau Distribution Function At The Edge}\label{appendixa}

This appendix details the derivation of the low-damping solutions in the lower-half complex plane under the limit of 
\(n_p \gg 1\).

For a low-damping solution, the analytic continuation of the dispersion relation into the lower-half plane, obtained using the Landau contour, is given by:
\begin{equation}
k^2 - \left[ \text{P.V.} \int_{-\infty}^{\infty} \frac{\partial f_p(v) / \partial v}{v - p/k} , dv + 2 i\pi \frac{\partial f_p}{\partial v} \bigg|_{v=p/k} \right] = 0
\label{a1}
\end{equation}

The distribution function \(f_p\) used in this work is not an analytic function over the entire complex plane.
 Instead, it is a meromorphic function featuring  \(n_p/2\)
 poles in both the upper and lower half-planes. These poles, denoted by \(v_{p0}\)
, are located at:
\begin{equation}
v_{p0}=v_0+\Delta v_p \exp \left(i \frac{2n-1 }{n_p} \pi \right) \quad n = 1,2 \cdots n_p
\end{equation}

For such a zeroth-order distribution function, the result from the 
Landau contour method remains valid, provided that the perturbed distribution 
function is analytic. \cite{stucchi2025landau} Furthermore, since the 
dispersion function appears in the denominator of the integrand for the 
inverse Laplace transform, the singularities of \(f_p\) do not pose any issue for the residue theorem. 
 However, this does not imply they are inconsequential to the evolution of the electric field. On the contrary, 
 these poles are crucial for the emergence of the zeros discussed in Sec \ref{section:2}. 
 A detailed calculation reveals that two accompanying zeros of the dispersion relation arise in the vicinity of each pole, 
 and these zeros correspond to the low-damping solutions of interest.

First, we assume that \(v_0\) is sufficiently large, such that when 
performing calculations in the vicinity of one plateau, the influence 
from the other plateau can be neglected. Under this assumption, the first 
derivative of \(f_p\) with respect to \(z\)
near the plateau is:
\begin{equation}
  \frac{\mathrm{d} f}{\mathrm{d} z}=  \frac{\mathrm{d} f_M}{\mathrm{d} z} -\frac{ \mathrm{d} f_M/\mathrm{d} z}{1+\left[(v-v_0)/\Delta v_p\right]^{n_p}}+
  \frac{f_M-f_M(v_0)}{\left\{1+\left[(v-v_0)/\Delta v_p\right]^{n_p}\right\}^2}\left[n_p \frac{(v-v_0)^{n_p-1}}{(\Delta v_p)^{n_p}}\right]
  \label{A3}
\end{equation}
where \(f_M(v_0)\)
 denotes the value of the Maxwellian distribution at \(v_0\)
, and all other parameters are defined as in Sec \ref{section:2}.

Next, we select a specific pole by fixing \(n\) and seek a solution of the form
 \(z_p=v_{p0}+ R \exp\left(i \theta \right)\), assuming \( R\ll  \Delta v_p \),
 Substituting this into Eq. \ref{A3} and expanding in terms of the small 
 parameter \(\epsilon_R = R/\Delta v_p\), we find that the three terms 
 on the right-hand side of Eq. \ref{A3} are of the order
\(O(1)\),\(O(1/\epsilon_R)\) and \(O(1/\epsilon_R^2)\), respectively.
By retaining terms up to \(O(1/\epsilon_R)\)  and discarding the unphysical solution
\(R=0\), we obtain a solution for \(R\)
 and \(\theta\):
 \begin{equation}
 \begin{aligned}
    R&=\frac{\Delta v_p }{n_p-1}\\
    \theta&=\frac{2n-1 }{n_p }\pi -\pi
 \end{aligned} 
 \label{A4}
\end{equation}

This solution is consistent with our initial assumption \(R\ll  \Delta v_p\)
in the limit \(n_p \gg 1\). Geometrically, this result implies that the points 
\(v_0 ,v_{p0}\) and \(z_{p}\) are collinear in the complex plane,  with \(z_p\)
situated between \(v_0\) and \( v_{p0} \).

By retaining terms up to the \(O(1)\) order, and letting \(x=(R/\Delta v_p)\exp \left[-i(2n-1)\pi/n_p +i\theta\right]\)
and \(\Delta f_M= f_M(v_{p0})-f_M(v_0)\), we obtain a quadratic equation with complex coefficients, \(A x^2+B x+C=0\)
( Note that the term \(x/R\) is a constant), the coefficient is expressed by the following equation.

\begin{equation}
  \begin{aligned}
    A&=Z_M n_p^2-\frac{n_p(n_p-1)}{2}\left.\frac{\mathrm{d} f_M}{\mathrm{d} z}\right|_{z=v_{p0}}-\frac{n_p^2(n_p-1)x}{2x}\Delta f_M \exp(-i\theta)\\
      &+\left.\frac{\mathrm{d^2} f_M}{\mathrm{d} z^2}\right|_{z=v_{p0}}\Delta v_p \exp \left(i \frac{2n-1 }{n_p} \pi - i\theta  \right)\\
    B&= -\Delta f_M \frac{n_p(n_p-1)x }{R}\exp(-i\theta)\\
    C&=-\Delta f_M \frac{n_p x }{R }\exp(i\theta) \\
    Z_M &=  \frac{1}{2i\pi} \left[ \int_{-\infty}^{\infty} \frac{\partial f_0(v) / \partial v}{v - p/k} \, dv +  2 i  \pi \frac{\partial f_M}{\partial v} (p/k) -k^2 \right]
  \end{aligned}
\end{equation}
As a result, the single solution splits into two, which is in 
qualitative agreement with the results presented in Sec \ref{section:2}.

\begin{figure}[h]
  \centering
  \includegraphics[width=\textwidth]{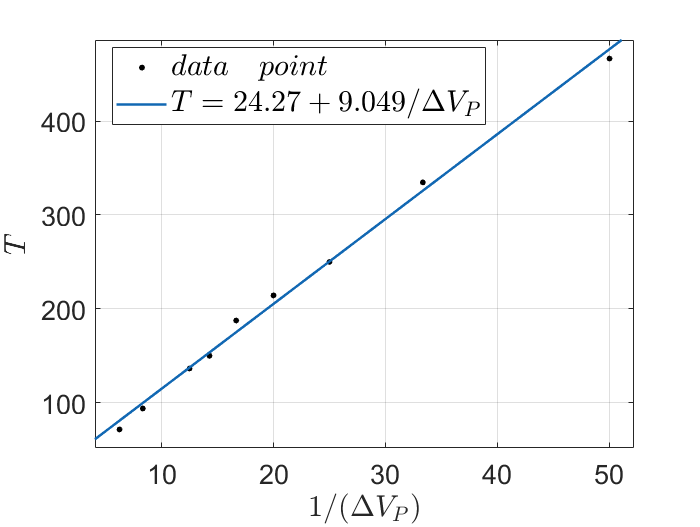}
  \caption{As the size of the plateau changes, the beat frequency period changes.}
  \label{ap1}
\end{figure}

If the correction term \(R\)
 is neglected, the two solutions merge into one, and the resulting beating period is given by 
 \(T=\pi/(\Delta v_p \cdot k)\). Numerical calculations of the beating 
 period for varying plateau widths are presented in Figure \ref{ap1}. 
 These results show that while the slope of the period versus the 
 inverse of the plateau width is not exactly \(\pi/k\), the 
 inverse relationship is well-maintained. The numerically obtained 
 slope is slightly larger than the value predicted by the simplified 
 model, a finding that is qualitatively consistent with our first-order 
 approximation, which places the low-damping solution between 
\(v_0\) and \( v_{p0} \).

\section{Symmetry of the Vlasov Equation}\label{appendixb}
For the Vlasov-Poisson system, under symmetric initial conditions:

\begin{equation}
  f(x,v,t=0)=f(L-x,-v,t=0)
\end{equation}
where \(L\)  represents the length of the one-dimensional system (which becomes the period length under periodic boundary conditions).

It can be proven that \(f(x,v,t)=f(L-x,-v,t)\), meaning that if the distribution function is centrally symmetric in phase space at the initial moment, this property remains unchanged as the Vlasov-Poisson system evolves over time.

First, we need to prove that if  \(f(x,v,t)\) and \(E(x,t)\) are solutions to the Vlasov-Poisson system, then 
\(f(L-x,-v,t)\) and \(-E(L-x,t)\)  are also solutions to the Vlasov-Poisson system.

Define the functions  \(f^\prime=f(L-x,-v,t)\) and \(E^\prime=-E(L-x,t)\), and substitute them into the Vlasov-Poisson equations,
we can find that:

\begin{equation}
  \dfrac{\partial  f^\prime}{\partial t}+v\dfrac{\partial f^\prime}{\partial x }-E^\prime \dfrac{\partial f^\prime}{\partial v }=0
\end{equation}
\begin{equation}
  \dfrac{\partial E^\prime}{\partial x}=1-\int f^\prime dv
\end{equation}

Since \( f(x,v,t)\) and \( E(x,t)\) are solutions to the Vlasov-Poisson system, the above equations are satisfied. Therefore, \(f^\prime(x,v,t  )\) and 
\( E^\prime(x,t   )\) are also solutions to the Vlasov-Poisson system.

Next, we need to prove that for the same initial conditions, the Vlasov-Poisson system has a unique solution, thereby completing the proof of this symmetry.

There are already well-established methods for proving the uniqueness of solutions for the Vlasov-Poisson system.  \cite{glassey1996cauchy_ch3,schaeffer1991global} 
The following briefly outlines the proof approach:

Assume that for the same initial conditions, the Vlasov-Poisson system has two inconsistent sets of solutions:
,\(f_1,E_1\) and \(f_2,E_2\). Denote the differences between the two 
sets of solutions as \(\delta f\) and \(\delta E\). By taking the difference of the 
Vlasov and Poisson equations for the two sets of solutions, we obtain:

\begin{equation}
  \frac{\partial (\delta f)}{\partial t} + v \frac{\partial (\delta f)}{\partial x} - E_1 \frac{\partial (\delta f)}{\partial v} = \delta E \frac{\partial f_2}{\partial v}
\end{equation}
\begin{equation}
  \frac{\partial (\delta E)}{\partial x} = - \int \delta f \, dv
\end{equation}

Multiply both sides of the Vlasov equation by \(\delta f\) and integrate over the entire phase space:

\begin{equation}
  \frac{1}{2} \frac{d}{dt} \iint |\delta f|^2 \, dx \, dv = \iint \delta f \cdot \delta E \cdot \frac{\partial f_2}{\partial v} \, dx \, dv
\end{equation}

To continue the proof, assume that \(\delta f\) has sufficiently good 
properties, and based on the Poisson equation being a linear equation in \(\delta f\), we have:
\begin{equation}
  \int \delta E \cdot \delta f \,dx\,dv <C \int(\delta f)^2 \, dx \, dv
\end{equation}
where \(C \) is a finite constant.

Let \(\epsilon_f(t) =  \int |\delta f(t)|^2 \, dx \, dv\), then we have:

\begin{equation}
  \frac{d\epsilon_f}{dt} \leq 2K\cdot \epsilon_f(t)
  \label{b8}
\end{equation}
where \(K\) is a constant function satisfying \(K\geqslant C\cdot \left(\frac{\partial f_2 }{\partial v}\right)_{max} \). Here, 
\(\left(\frac{\partial f_2 }{\partial v}\right)_{max}\) represents the maximum value of \(\frac{\partial f_2 }{\partial v}\) during the time evolution. 
Since \(\frac{\partial f_2 }{\partial v}\) is finite, there always exists a constant \(K\) that satisfies the above inequality.

Given that the initial conditions are the same, \(\epsilon_f(t=0)=0\). According to 
the above inequality, we obtain \(\epsilon_f(t)=0\), which implies \(\delta f=0\). Therefore, the Vlasov-Poisson 
system has a unique solution for the same initial conditions.

For general symplectic structure algorithms,  \cite{cheng1976integration} 
if the initial perturbation is symmetric, the symmetry of the 
distribution function in phase space cannot be guaranteed during the 
time-stepping process. To address this, the algorithm used in this study 
calculates only half of the phase space distribution function at each 
time step, with the other half directly given by symmetry. This approach 
not only reduces the computational load by nearly half (since the 
distribution function propagation, which utilizes cubic spline 
interpolation, is the most computationally intensive part of the program) 
but also ensures that the symmetry remains unchanged over time.

Without choosing a time-stepping method that preserves symmetry, 
using a conventional symplectic algorithm to calculate the steady 
state of large-amplitude evolution can lead to strong distortions of 
the distribution function near zero velocity, failing to maintain 
symmetry and resulting in non-physical solutions. This phenomenon only 
occurs in large-amplitude evolutions; for small-amplitude evolutions, 
the destruction of symmetry in the steady state is not significant.

To demonstrate the necessity of the symmetry-preserving time-stepping method, 
simulations were conducted with 
\(\beta=0.3\), \(N_x=2048\) and \(N_v=16384\) for both the general 
symplectic algorithm and the symmetry-preserving algorithm. 
The distribution functions at  
\(t=3750\omega_p^{-1}\) are shown in the figure below.

Figure \ref{appendixb1} shows the distribution function calculated by symmetric stepping, while Figure \ref{appendixb2} presents the distribution function 
obtained from asymmetric calculation. It can be observed that the symmetry is no longer evident in Figure \ref{appendixb2}, 
which leads to differences in the shape, quantity of vortices as well as their positions in the phase space between the two figures.

\section{Verification of the Numerical Simulation}\label{appendixc}

To validate the reliability of the preceding physical results, we examine the conservation of the invariants 
\(I_3\) and the entropy \(S\). These quantities are defined as

\begin{equation}
  I_3=\int f^3 \mathrm{d}x \mathrm{d}v
\end{equation}
\begin{equation}
  S=\int f\ln f \mathrm{d}x \mathrm{d}v
\end{equation}

\added{First, to exclude the influence of numerical resolution on the physical results, and in 
particular to verify the authenticity of the fine phase-space structures 
observed in Section \ref{section:4}, we investigate the convergence of the simulation 
results with respect to grid resolution. Since the convergence regarding 
the velocity grid number \(N_v\) and its impact on filamentation has been extensively studied in previous works,
\cite{carril2023formation} we focus here on the influence of the spatial resolution \(N_x\).}

\begin{figure}[h]
    \centering 
    \includegraphics[width=\textwidth]{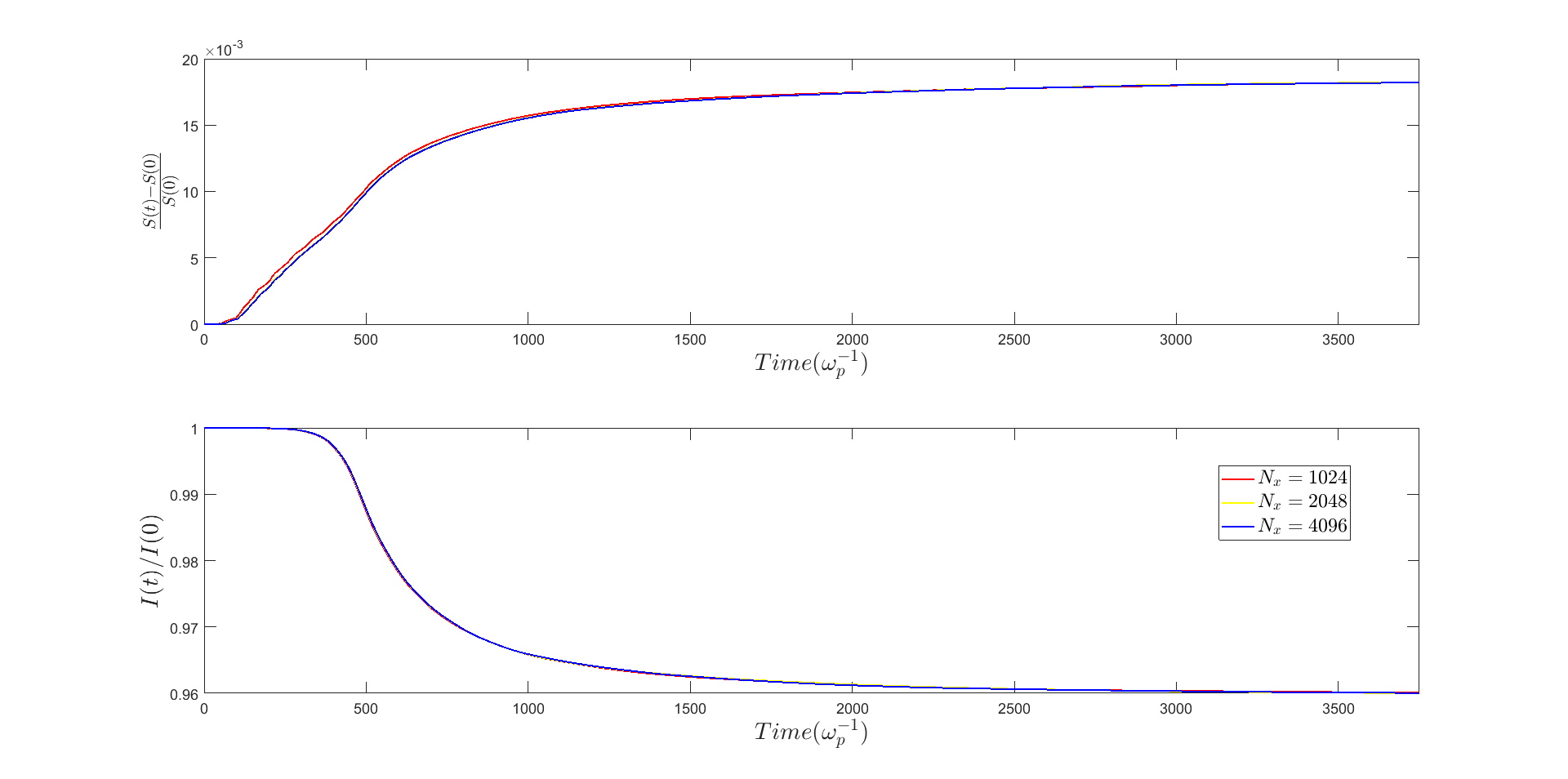}
    \captionsetup{justification=raggedright,singlelinecheck=false}
    \caption{Time evolution of  the relative entropy change \((S(t)-S(0)/S(0))\) 
     and  the third-order invariant \(I_3(t)/I_3(0)\) for varying spatial grid numbers
     \(N_x\). The simulation parameters are set to \(\beta=0.15\) and \(N_v=16384\).}
    \label{appendix_c1}
\end{figure}

\added{Figure \ref{appendix_c1} illustrates the evolution of the invariants \(S\) and \(I_3\) under
different spatial grid numbers \(N_x= 1024 ,2048,4096\), with a fixed initial perturbation amplitude
\(\beta=0.15\) and velocity grid \(N_v=16384\). The curves for the three resolutions overlap almost perfectly, indicating that
\(N_x=2048\) is sufficient to ensure the numerical convergence of global physical quantities.}

\begin{figure*}[t]
    \centering 
     \captionsetup[subfigure]{labelformat=empty}
    \begin{subfigure}[t]{0.8\textwidth} 
        \centering
        \includegraphics[width=\linewidth,height=5.5cm]{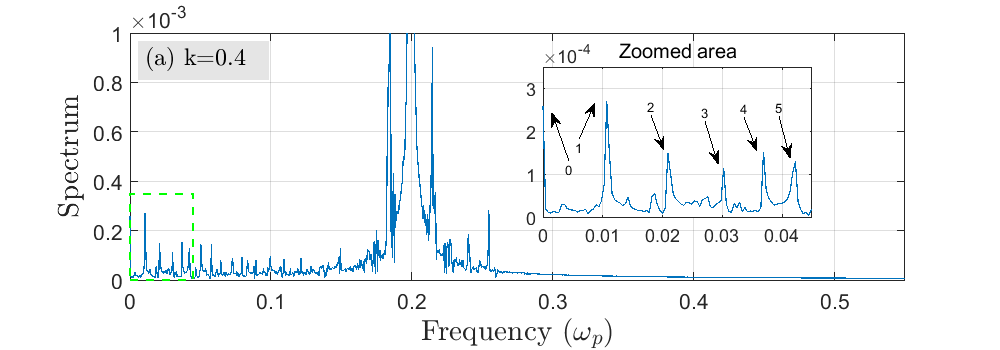} 
        \caption{}
        \label{fig:7a}
    \end{subfigure}%
    \\
    \begin{subfigure}[t]{0.8\textwidth} 
        \centering
        \includegraphics[width=\linewidth,height=5.5cm]{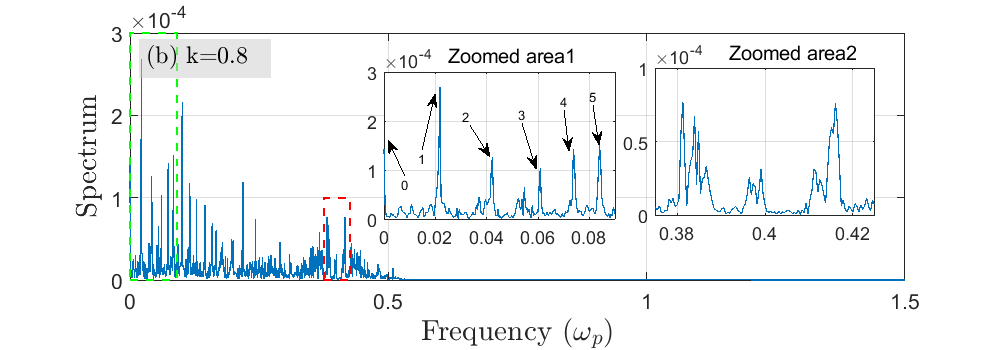} 
        \caption{}
        \label{fig:7b}
    \end{subfigure}%
    \\
    \begin{subfigure}[t]{0.8\textwidth}
      \centering
        \includegraphics[width=\linewidth,height=5.5cm]{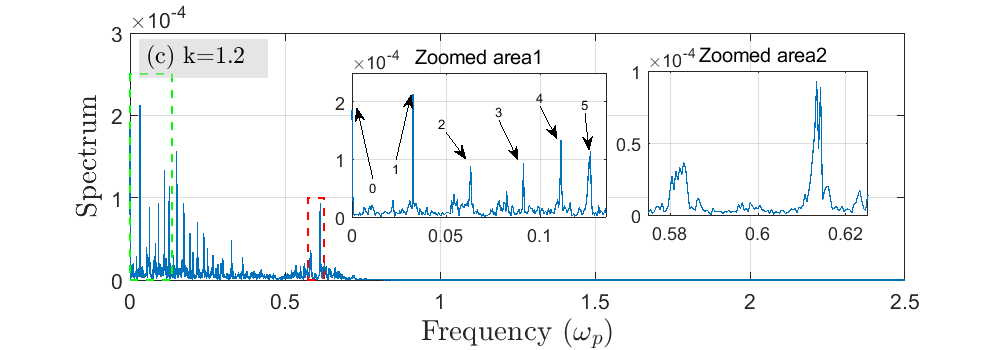}
        \caption{}
        \label{fig:7c}
    \end{subfigure}%
    \captionsetup{justification=raggedright,singlelinecheck=false}
    \caption{Fourier components of the electric field at different wavenumbers \(k_E\) at \(\beta=0.15,k=0.4\) and \(t=2500\sim 3750\),
    (a) \(k_E=0.4\),  (b) \(k_E=0.8\),  (c) \(k_E=1.2\).} 
    \label{fig:main7}
  \end{figure*}

\begin{figure}[h]
    \centering 
    \includegraphics[width=\textwidth]{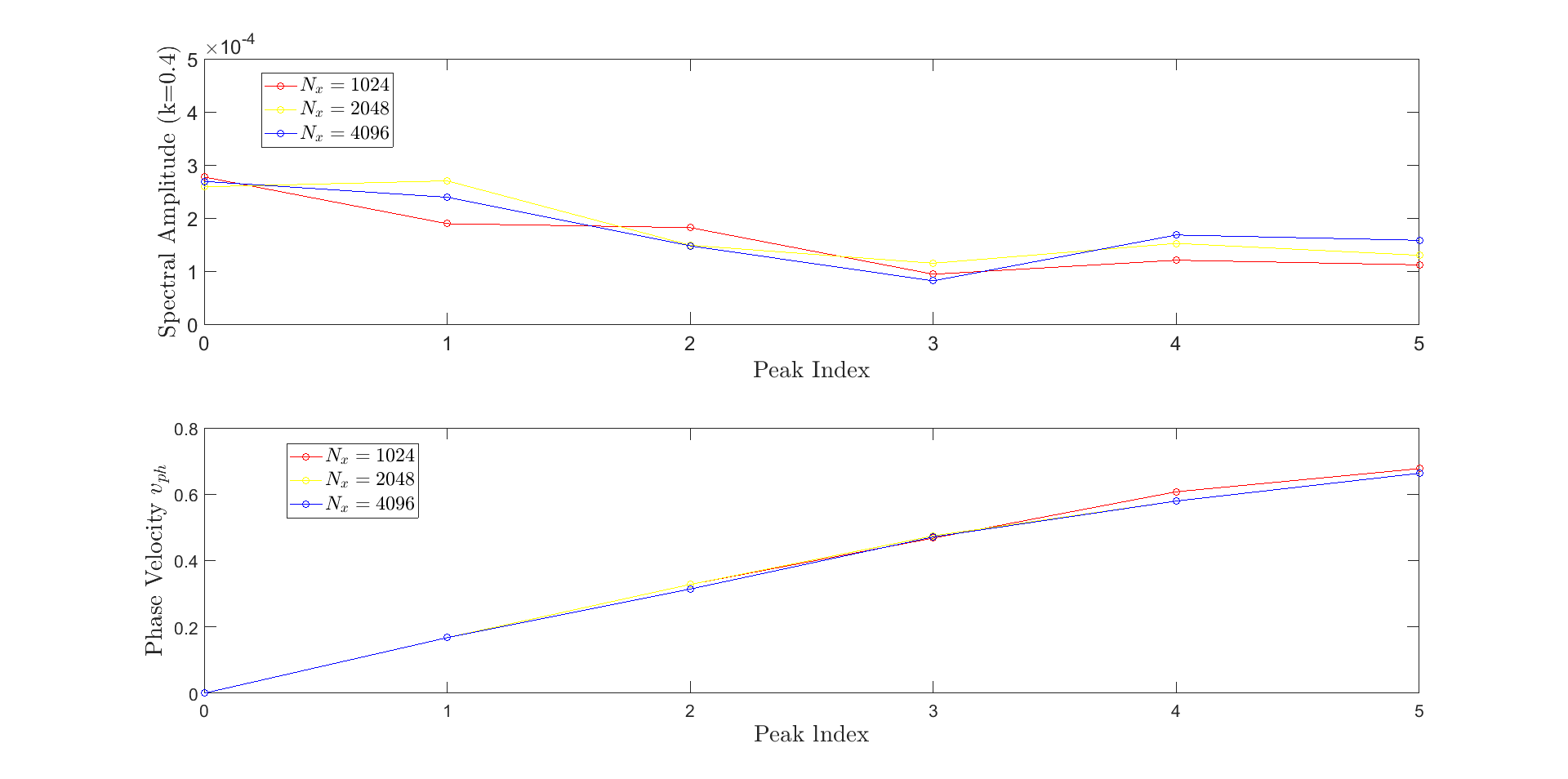}
    \captionsetup{justification=raggedright,singlelinecheck=false}
    \captionsetup{justification=raggedright,singlelinecheck=false}
    \caption{Convergence analysis of the low-frequency characteristic peaks (labeled 0-5 in Fig. \ref{fig:7a}.)
    with respect to the spatial grid number \(N_x\). The top panel 
    displays the spectral intensity of each peak, while the bottom 
    panel shows the corresponding phase velocity. }
    \label{appendix_c2}
\end{figure}

\added{Moreover, to confirm the physical validity of the vortex structures in the low phase-velocity region identified in
Figs.\ref{fig:main7} and \ref{fig:main8}, we examined the variation of their spectral characteristics with spatial resolution.
Figure \ref{appendix_c2} displays the convergence of the spectral amplitude (top panel) and the corresponding phase velocity (bottom panel) 
for the low-frequency discrete peaks (labeled 0-5 in Fig. \ref{fig:7a}) across various \(N_x\) numbers. In  contrast to the numerical artifacts at
\(v_p/n \) discussed in Section \ref{section:4}, which vanish as resolution increases, the intensity and location of these 
low-frequency peaks remain  stable as \(N_x\) increases. This robustness against spatial resolution  demonstrates that these 
low-velocity vortices are intrinsic physical structures of the plasma asymptotic state, rather than artifacts arising from numerical discretization.
}

\begin{figure}[h]
  \centering
  \captionsetup[subfigure]{labelformat=empty}
    \begin{subfigure}[t]{0.8\textwidth}
        \centering
        \includegraphics[width=0.65\textwidth]{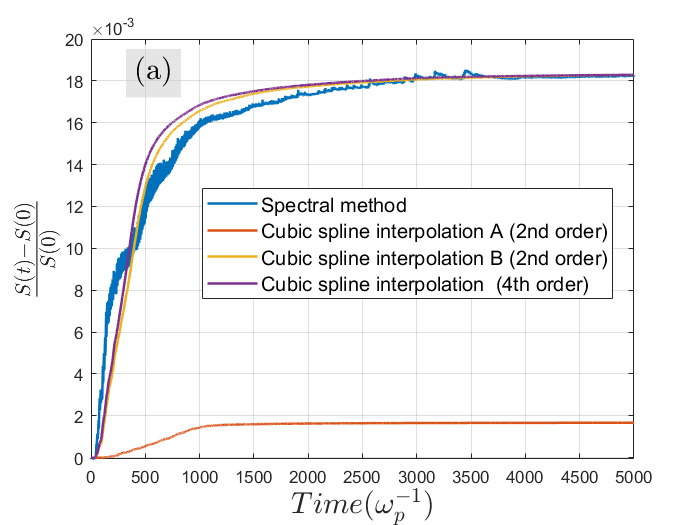}
        \caption{}
        \label{appendixc3}
    \end{subfigure}\\
    \begin{subfigure}[t]{0.8\textwidth}
        \centering
        \includegraphics[width=0.65\textwidth]{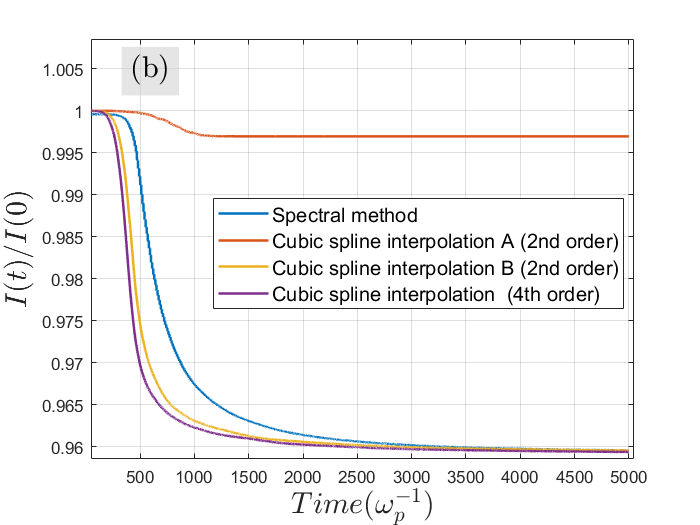}
        \caption{}
        \label{appendixc4}
    \end{subfigure}
    \captionsetup{justification=raggedright,singlelinecheck=false}
  \caption{Time evolution of (a) the entropy \(S\) and (b) the invariant \(I_3\) of simulations run with different numerical calculation methods and perturbation amplitude at \(N_x=2048\),\(N_v=8192\).
 Case 'A' corresponds to an initial amplitude of \(\beta = 0.04\), while case 'B' and the spectral/4th-order methods use \(\beta = 0.15\).}
  \label{appendix_c} 
\end{figure}

\added{Having established the convergence of spatial resolution, 
we further characterize the influence of numerical dissipation on the simulation 
by monitoring the evolution of these two invariants over time.}
Figure \ref{appendix_c} illustrates the impact of different numerical 
methods and initial perturbation amplitudes on these quantities. 
Three distinct numerical methods were employed for this comparison: 
the second-order symplectic  algorithm with cubic spline 
interpolation used in Sec \ref{section:3}, a fourth-order symplectic 
integrator, \cite{watanabe2005vlasov} and a second-order symplectic 
integrator combined with a spectral method that utilizes a high-order 
exponential filter. \cite{cheng1976integration,izrar1989integration,carril2023formation}

For a simulation with an initial amplitude \(\beta=0.15\), both \((S(t)-S(0))/S(0)\)
and \(I_3(t)/I_3(0)\) converge to the same final values across all three methods, yielding
\((S(t)-S(0))/S(0)=0.0182\) and \(I_3(t)/I_3(0)=0.96\). Regarding the entropy evolution, the 
cubic spline interpolation performs better in the early stage, whereas the spectral method 
is superior in the later stage. For the evolution of \(I_3\), the spectral method consistently 
outperforms the cubic spline interpolation. Contrary to expectations, the fourth-order 
symplectic integrator did not mitigate these numerical effects; instead, it accelerated the 
deviation of the invariants in the early stages of the simulation. This is likely attributable 
to the accumulation of numerical errors over the multiple steps inherent in the higher-order 
method. This outcome indicates that the time step size is not the primary source of this 
numerical phenomenon.

The magnitude of this numerical effect is strongly dependent on the initial amplitude. 
When the amplitude is reduced from 0.15 to 0.04, the effect is significantly attenuated. 
The final converged values change to \((S(t)-S(0))/S(0)=0.0017\) and \(I_3(t)/I_3(0)=0.997\).
Compared to the large-amplitude case, the numerical deviation is reduced by approximately an 
order of magnitude, demonstrating that the phenomenon intensifies with increasing initial 
perturbation.

\begin{figure}[h]
    \centering 
    \includegraphics[width=\textwidth]{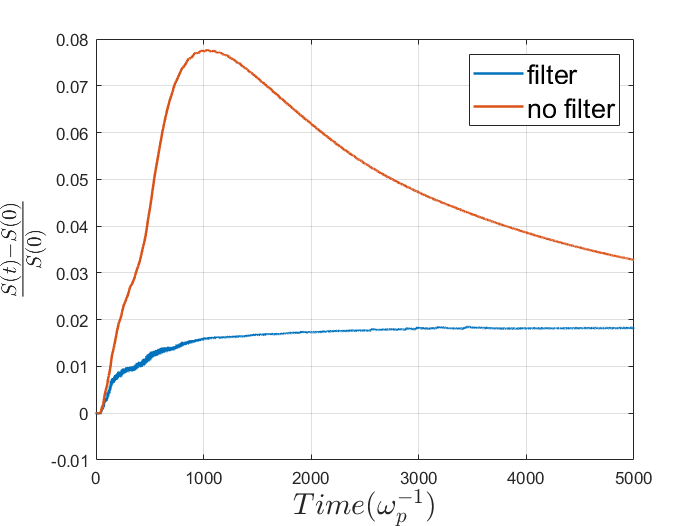}
    \captionsetup{justification=raggedright,singlelinecheck=false}
    \caption{ Comparison of entropy evolution in Vlasov simulations with and without filtering. Here \(\beta=0.15 ,N_x=2048,N_v=8192\).}
    \label{appendix_c5}
\end{figure}

Furthermore, when using the spectral method, the entropy exhibits an oscillatory growth, 
as shown in Figure \ref{appendix_c}. The instances of entropy decrease coincide precisely 
with the application of the filter. This suggests that the numerical artifact originates from 
the filamentation of the distribution function, and the suppression of this filamentation by 
the filter alleviates the non-conservation. To substantiate this claim, we performed a 
simulation using the spectral method without any filtering (see Fig. \ref{appendix_c5}) and 
observed a dramatic enhancement of the numerical artifacts. The inherently low numerical 
dissipation of the spectral method (which decreases exponentially with \(N_v\))
 is incapable of suppressing filamentation, this results in more pronounced 
 numerical artifacts than those observed with the highly dissipative cubic spline 
 interpolation (where dissipation is of order \(O(N_v^{-n-1})\)), thereby highlighting 
 the necessity of filtering in spectral method simulations.

\section*{DATA AVAILABILITY}
The data that support the findings of this study are available from
the corresponding author upon reasonable request.

\nocite{*}
\bibliographystyle{aipnum4-1}
\bibliography{apssamp}

\clearpage

\newpage

\clearpage

\newpage
\newpage
\newpage
  \newpage
\clearpage

\newpage
\newpage
\newpage
\begin{figure}[h]
  \centering
  \captionsetup[subfigure]{labelformat=empty}
    \begin{subfigure}[t]{0.8\textwidth}
        \centering
        \includegraphics[width=0.65\textwidth]{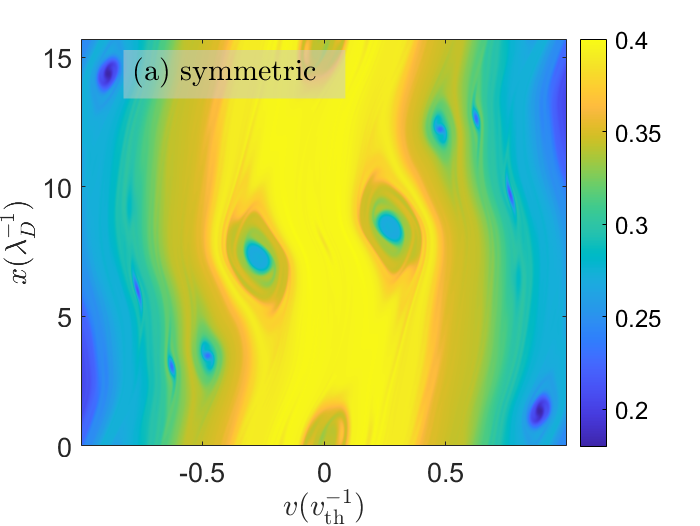}
        \caption{}
        \label{appendixb1}
    \end{subfigure}\\
    \begin{subfigure}[t]{0.8\textwidth}
        \centering
        \includegraphics[width=0.65\textwidth]{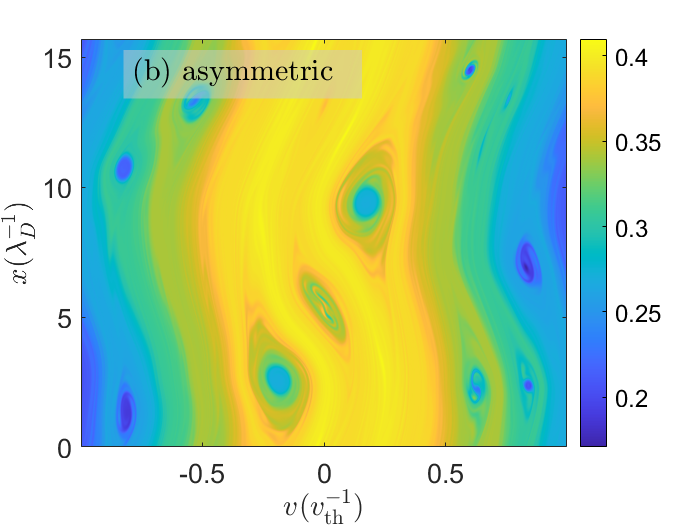}
        \caption{}
        \label{appendixb2}
    \end{subfigure}
    \captionsetup{justification=raggedright,singlelinecheck=false}
  \caption{Distribution functions obtained by two different stepping methods, (a) and (b) show 
  the progressive phase space structure obtained by using symmetry format and without symmetry separation under symmetric perturbations, respectively.}
  \label{appendix_b} 
\end{figure}

\end{document}